\documentclass[12pt]{elsarticle}
\usepackage{amssymb,amsmath,bm,mathtools,xcolor,float}
\numberwithin{equation}{section}
\usepackage[ruled,linesnumbered,vlined]{algorithm2e}
\usepackage{graphicx}
\usepackage[colorlinks=true,linkcolor=blue]{hyperref}
\usepackage{geometry}
\usepackage{subcaption}	
\usepackage{lineno}
\begin{document}

\title{Meso-scale size effects of material heterogeneities on crack propagation in brittle solids: Perspectives from phase-field simulations}

\author[1]{Liuchi Li\corref{cor1}}
\ead{lli128@jhu.edu}
\cortext[cor1]{Corresponding author} 
\author[2]{Jack Rao}
\author[1,2,3]{Todd Hufnagel}
\author[1,2,3]{KT Ramesh}

\address[1]{Hopkins Extreme Materials Institute, Johns Hopkins University, Baltimore, MD 21218, USA}
\address[2]{Department of Mechanical Engineering, Johns Hopkins University, Baltimore, MD 21218, USA}
\address[3]{Department of Materials Science and Engineering, Johns Hopkins University, Baltimore, MD 21218, USA}

\begin{abstract}
Brittle solids are often toughened by adding a second-phase material. This practice often results in composites with material heterogeneities on the meso scale: large compared to the scale of the fracture process zone but small compared to that of the application. The specific configuration (both geometrical and mechanical) of this mesoscale heterogeneity is generally recognized as important in controlling crack propagation behavior and, subsequently, the (effective) toughness of the composite.
Here, we systematically investigate how dynamic brittle fracture navigates through a linear array of mesoscale inclusions. Using a variational phase-field (PF) approach, we compute the apparent crack speed and fracture energy dissipation rate to compare crack propagation (and the resulting toughening) under Mode-I loading for various configurations of inclusions. We identify an interplay between the size of inclusion and that of the $K$-dominant zone in the presence of elastic heterogeneity: matching these two sizes gives rise to the best toughening outcome for a given area fraction of inclusions. We discuss mechanisms that rationalize this observation and the importance of the length scale parameter used in PF models in interpreting simulation results. Our work sheds physical insight into the interaction between size effects and material properties, thereby opening a venue for the rational design of functional (architected) composites for dynamic fracture applications.

\end{abstract}
 \maketitle

\section{Introduction}

Structural materials such as glass and ceramics find applications across multiple industrial sectors, from aerospace to defense, where extreme environments (such as high-speed impact) are often encountered. However, these strong and lightweight materials are often brittle. Thus, it has been a common practice to toughen a brittle material by adding a second material phase to alter the fracture behavior. Such approaches often result in mesoscale material heterogeneities: the heterogeneity length scale is large compared to that of the process zone, but small compared to that of the application (which gives rise to ``extrinsic toughening mechanisms" \cite{ritchie2011conflicts, hossain2014effective}). As such, this practice enjoys a very high-dimensional design space in terms of not only what material to choose as the second phase but also how to configure this second phase spatially with respect to the base material phase, given the significant separation of length scales.

The biggest challenge, therefore, lies in how we effectively explore this design space (in terms of choosing materials and their geometrical configurations) to minimize the well-known tradeoff between strength and toughness \cite{ritchie2011conflicts, ashby2013materials}. For instance, adding a more compliant or tougher phase can help arrest a propagating crack, thereby toughening the material. However, such inclusions can also lead to a decrease in the overall stiffness and strength. Historically, this design space has been explored by focusing on the choice of the second-phase material. Examples include crystallized ceramic inclusions in an LS$_2$ glass matrix \cite{serbena2015crystallization} and TiC particles in a SiC ceramic matrix \cite{wei1984improvements}. The issue is that, partly due to synthesis and processing limitations, the resulting geometrical configuration of the material heterogeneity is usually stochastic and not well controlled, leaving the geometrical aspect (a considerable portion of this design space) largely unexplored. 

Recently, advances in manufacturing techniques allow the precise control of a material's structure across scales, making it possible to systematically explore the geometric aspect of this design space. There are particularly promising opportunities within the context of mechanical ``meta-materials"\footnote{The terms ``metamaterials" and ``architected materials" are often used interchangeably in the literature.} which have already demonstrated novel mechanical properties (such as a high stiffness-to-density ratio \cite{zheng2014ultralight,meza2014strong}, with chiral character \cite{frenzel2017three}, and being (re-)programmable \cite{wang2021structured,chen2021reprogrammable,liu2022growth}) that are not found in conventionally manufactured materials. \citep{kadic20193d, xia2022responsive} provide comprehensive introductions to this topic. An emerging research area is that of studying the resistance to fracture of these metamaterials \cite{shaikeea2022toughness, da2022data, jia2023controlling, karapiperis2023prediction, magrini2023control}, particularly the class of metamaterials containing arrays of compliant inclusions (which often take the form of voids \cite{heide2020mechanics, fulco2022decoupling, liu2020high}). These studies recognized that the geometrical configuration of these inclusions (mainly size and spacing) could alter crack propagation behaviors \cite{hossain2014effective, bleyer2017dynamic, heide2020mechanics}. It has been argued that the K-dominant zone matters in this regard \cite{hossain2014effective, brodnik2021fracture}. A properly designed configuration can lead to high toughness \cite{liu2020high, conway2021increasing} and even directional asymmetry in toughness when designing inclusion shape comes into play as well \cite{brodnik2021fracture}.

Fracture propagation in a brittle composite with mesoscale heterogeneities can be highly dynamic and is known to show considerable rate effects \cite{albertini2021effective}, even if the remote macroscopic loading is quasi-static \cite{lazzaroni2012role}. However, a systematic and quantitative understanding is still lacking regarding how dynamic brittle fracture navigates through mesoscale heterogeneities, and what is the resulting toughening outcome \cite{michel2022merits}. Many interesting questions remain unanswered, especially from a physics-guided design perspective. For instance, why should we choose a specific inclusion size over another? Does changing the inclusion size necessitate a change of inclusion spacing to maintain a ``sweet spot" design? Lastly, how important are the material properties of the inclusions? These questions are also relevant for minimizing the undesired effects of these inclusions (for instance, those in the form of voids) on the overall material stiffness and strength, which are especially important for applications in extreme environments.

Here, we systematically quantify the mesoscale size effects of material heterogeneities on dynamic fracture propagation in a brittle composite (or meta-) material, using a dynamic phase-field approach for numerical simulation. This approach enables us to calculate fracture energy dissipation rate and crack speed at every instant. We consider inclusions with toughness and elastic contrast as mesoscale material heterogeneities, and we model them and the base medium as separate continuum solids that homogenize out any possible micro-scale (and below) material heterogeneities. In particular, we consider inclusion arrangements that go from being large but sparse on one end to being small but dense on the other end while holding the area fraction constant. We are particularly interested in comparing crack propagation and the associated toughening outcome of these configurations with inclusions possessing different spatial arrangements but identical area fractions. The underlying physical interpretations can have practical implications in designing flaw-insensitive composites that are resistant to crack propagation or prevent a crack from growing indefinitely.

The remainder of this paper is organized as follows. In Section~\ref{pf}, we briefly introduce the variational phase-field approach for dynamic fracture simulation; in Section~\ref{pfmodel}, we discuss our numerical model, parameter selection, and the analysis procedure; in Section \ref{kzonewithl}, we discuss the estimation of the K-dominant zone in the context of phase-field simulations; in Section~\ref{result}, we discuss the simulation results that demonstrate the relative size interplay between the inclusion and the K-dominant zone; in Section~\ref{sum}, we present a summary of our work and discuss limitations and future directions.

\section{Variational phase-field approach to fracture}\label{pf}
Phase-field modeling provides a mathematical framework that is widely used to describe physical systems, especially those with evolving interfaces far from equilibrium (fracture propagation in solids is a typical example). Since its first introduction in the context of solidification and phase transition \cite{cahn1958free}, it has been adapted to modeling many other phenomena such as multiphase flow \cite{lowengrub1998quasi, fu2018nonequilibrium}, collection cell dynamics \cite{najem2016phase, monfared2023mechanical}, and material failures \cite{bourdin2000numerical, bourdin2008variational, akerson2023optimal}, to name a few. At the core of phase-field modeling is a mathematical description of the energy (density) of a physical system $\gamma_\ell$, a quantity associated with the particular physical field of modeling interest. This is done using a scalar field $\phi \in [0,1]$  that varies smoothly in space over a length scale parameter $\ell$. In the particular case of fracture, we can view $\gamma_\ell$ as a (regularized) fracture energy density over the entire simulation domain, with the length scale $\ell$ being used to effectively smear out the crack, so that $\phi = 0$ typically indicates intact material and $\phi = 1$ indicates completely damaged material. It has been shown that $\ell$ determines the threshold for crack nucleation \cite{bourdin2014morphogenesis, tanne2018crack}, and from a physics standpoint, it can be viewed as a representation of active mechanisms in the process zone \cite{janssen2004fracture}. When modeling dynamic fracture problems, we seek to minimize an incremental Lagrange energy functional $\mathbf{I}_{\ell}$ using the principle of least action (see \cite{miehe2010phase,borden2012phase} for a comprehensive discussion relevant to the topic). The functional form is as follows:

\begin{align}
\mathbf{I}_\ell(u,\dot{u},\phi) = \int_{t_1}^{t_2}\left\{  \int_{\Omega} \left[  \frac{\rho}{2}|\dot{u}|^2- \mathcal{W}^e(u,\phi)-G_
\text{C}\gamma_{\ell}(\phi,\nabla \phi) +\rho b\cdot u \right] \text{d$\Omega$}  +\int_{\partial\Omega}   t \cdot u \text{dS}    \right\}\text{dt},
\label{pfintegral}
\end{align}
under the constraint $\dot{\phi} > 0$, to account for the irreversibility of the fracture process. In Eqn. \ref{pfintegral}, $u$ is the displacement field, with $\dot{u} = \frac{\partial u}{\partial t}$ the velocity field, $\rho$ the material density, $\phi$ the phase field parameter indicating the degree of material damage, $\mathcal{W}^e$ the elastic strain energy density, $G_\text{C}$ the critical energy release rate (or fracture toughness) \cite{griffith1921vi}, $\gamma_\ell$ the (regularized) fracture energy density, $b$ the gravitational constant, and $t$ the surface traction. Under plane stress conditions, we also have $G_\text{C} = K_\text{IC}^2/E$, where $K_\text{IC}$ is the Mode-I critical stress intensity factor. The form (or degree of complexity) of $\mathcal{W}^e$, $\Gamma_\ell$, and $G_\text{C}$ may be problem-specific (e.g., $G_\text{C}$ may be anisotropic \cite{bleyer2018phase, gerasimov2022second} and $\gamma_\ell$ may be dependent on higher-order terms of $\phi$ \cite{amiri2016fourth} ) and even Eqn. \ref{pfintegral} can be modified to model ductile instead of brittle fracture \cite{ambati2015phase,choo2018coupled,brach2019phase}. In this work, we restrict ourselves to materials that are isotropic and linear elastic with rate-independent fracture toughness. We can describe a material by three parameters: Young's modulus $E$, Poisson's ratio $\nu$, and fracture toughness $G_\text{C}$. Interfacial effects \cite{ming1989crack} can be included by incorporating a cohesive zone model \cite{rezaei2021direction}, but we neglect them for the purposes of this work, leaving that to a subsequent effort. We adopt the following form of $\gamma_\ell$, commonly used for modeling brittle fracture \cite{miehe2010phase}:

\begin{align}
\gamma_\ell = \frac{1}{4c_w\ell}\left(w(\phi)+\ell^2|\nabla \phi|^2 \right), \text{with}\, c_w = \frac{1}{2}\,\,\text{which implies}\,\, w(\phi) = \phi^2.
\label{pfdensity}
\end{align}

Note that this form is known as the AT2 model \cite{ambrosio1990approximation}, which is also implemented in \cite{borden2012phase} but is different from the so-called AT1 model \cite{pham2011gradient} (i.e. $c_w = 2/3, w(\phi) = \phi$) used in \cite{bleyer2017dynamic}. Although formulated differently, these two models have been shown to give similar results in terms of modeling crack initiation and propagation (see \cite{kristensen2021assessment} for a detailed discussion). We implement this method using FEM in 2D based on our previous work on modeling multibody contact mechanics problems \cite{li2021emerging}. The source code is publicly available at {\href{https://github.com/liuchili/Variational-phase-field-method-for-dynamic-fracture-problem.git}{\color{blue}{https://github.com/liuchili/Variational-phase-field-method-for-dynamic-fracture-problem.git}}}. 

Our implementation finds the stationary solution to the minimization of Eqn. \ref{pfintegral} in parallel based on the alternating minimization scheme, utilizing an in-house conjugate gradient solver that runs with OpenMP and MPI. In particular, we use the average acceleration scheme to solve for $u$. At each time step, we iterate back and forth between $u$ and $\phi$ until convergence is achieved. We verified our implementation using the classical Kalthoff-Winkler experiment \cite{kalthoff2000modes} (see Appendix for details).

\section{Modeling fracture propagation in brittle solid with meso-scale heterogeneity}\label{pfmodel}
\subsection{Model setup and parameter selection}
\begin{figure}[h]
\centering
\includegraphics[width=\linewidth]{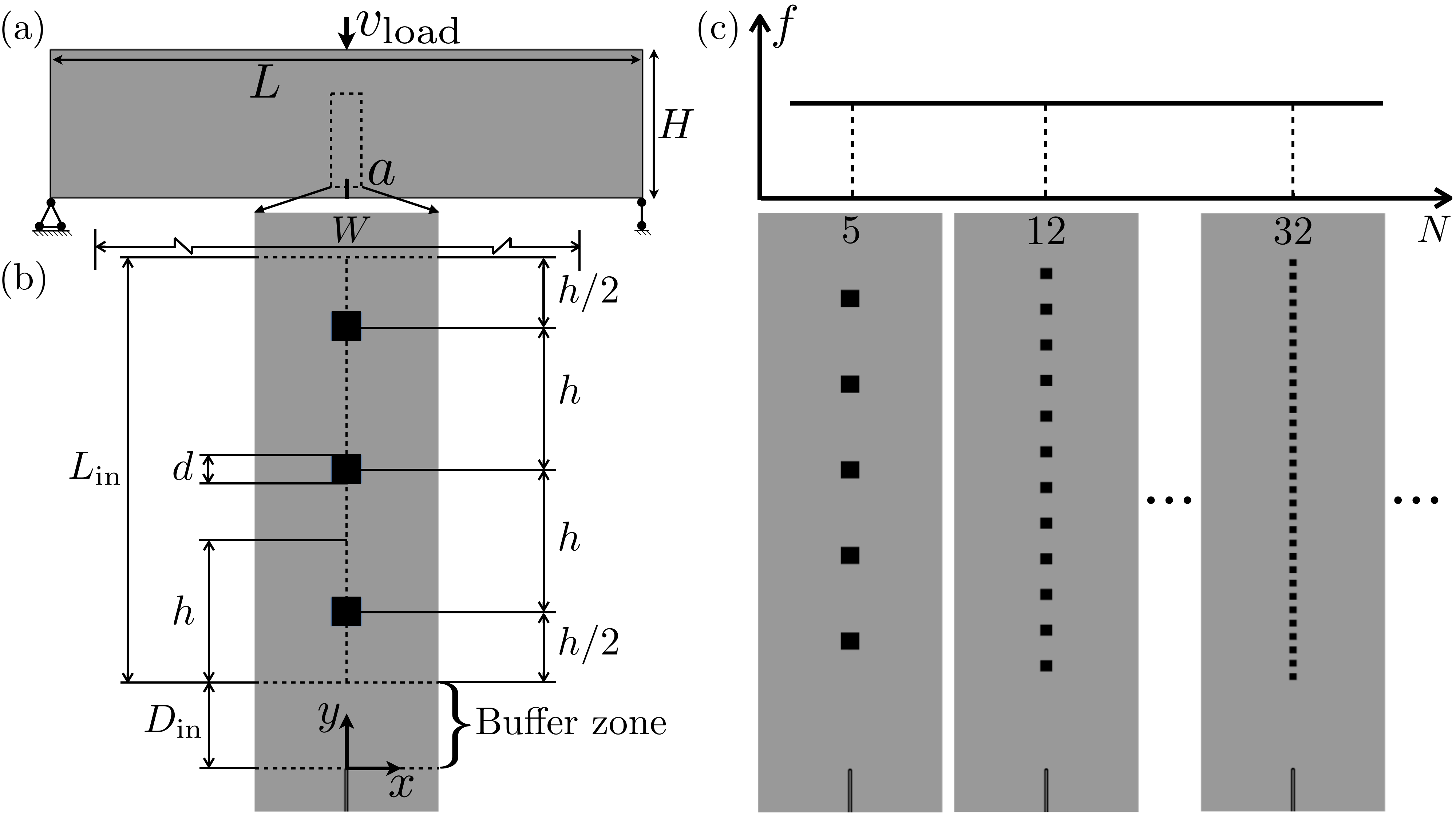}
\caption{(a) The setup of our numerical model: a single-notched three-point bending beam subjected to a constant indentation speed $v_\text{load}$. The beam has a span $L$, a height $H$, and a notch with length $a$. (b) The geometric configuration of a single array of mono-sized ($d$) and equally spaced ($h$) inclusions embedded along a line of length $L_\text{in}$ that starts at a distance of $D_\text{in}$ (as a buffer zone) from the notch tip. The width $W = 10L_\text{in}$ considered for computing the area fraction is not drawn to scale. In this particular image, $N = 3$ using $c_0 = 0.2, N_0 = 5$. (c) Three examples of the geometrical configuration produced with $c_0 = 0.2, N_0 =5$ for $N = 5$, $N = 12$, and $N = 32$.}
\label{config}
\end{figure}

We use the variational phase-field method discussed above to simulate dynamic crack propagation under Mode-I crack opening mode in plane stress condition, using a single-notched three-point bending configuration (with span $L$, height $H$, and notch length $a$) that is subjected to a constant indentation velocity $v_\text{load}$ as shown in Fig. \ref{config}(a). We consider a simple case where we represent mesoscale material heterogeneities as square inclusions. We arrange these inclusions (with a uniform size $d$) as a single array (with a uniform spacing $h$) along a line aligned with the expected crack propagation path, starting from a distance of $D_\text{in}$ from the notch tip (a ``buffer zone") and extending to a length of $L_\text{in}$, as shown in Fig. \ref{config}(b). Next, we introduce two design parameters, the number of inclusions $N$ and the relative spacing $c = d/h$. Since $h = L_\text{in}/N$, $f = \frac{Nd^2}{WL_\text{in}}$ can also be expressed as $f = \frac{c^2L_\text{in}}{WN}$. In this work, we fix $W = 10L_\text{in}$ (see Fig. \ref{config}(b)), which is large enough such that (1) the crack can only interact with a single array of inclusions, and (2) the stress fields developed from any two neighboring inclusion arrays (with a separation distance of $W = 10L_\text{in}$) do not interact\footnote{Note that this can be a conservative estimation.}. To keep the computational cost reasonable, in our numerical model we only consider a single array of inclusions directly above the notch. Plugging in $W = 10L_\text{in}$, we have $f = \frac{c^2}{10N}$. We use two baseline values (denoted as $c_0$ and $N_0$) to fix $f = \frac{c_0^2}{10N_0}$. Accordingly, we can get different $d$ by varying $N$ as follows:

\begin{align}
d = \frac{c_0L_\text{in}}{\sqrt{N_0N}}\,\,\text{with}\,\, h = \frac{L_\text{in}}{N}, \text{implying}\,\, c = \frac{d}{h} = c_0\sqrt{\frac{N}{N_0}}.
\label{identicalformula}
\end{align}



So, for a given value of $c_0$ and $N_0$, varying $N$ provides different geometrical configurations represented by different combinations of $(d,h)$. Fig. \ref{config}(c) shows three examples using this design strategy with $c_0 = 0.2, N_0 = 5$ for $N = 5$, $N = 12$, and $N = 32$, respectively. Finally, we choose $L = 32a, H = 8a, L_\text{in} = 5a$, and $D_\text{in} = a$, where $a$ is the initial crack length (notch length) chosen as $10$ mm in this work. We pick such a geometry to realize a large enough domain in the sense that the crack initiation and propagation are not affected before entering the inclusion-embedded region (see Fig. \ref{model} for a representative example). In this way, it becomes convenient to compare crack propagation for different inclusion configurations.

In addition to the geometrical configurations, we also vary the mechanical configurations by selecting different values for $\alpha$ (representing the elastic contrast) and $\beta$ (representing the toughness contrast):

\begin{align}
\alpha = E_\text{in}/E_0,\,\,\text{and}\,\, \beta = G_\text{C,in}/G_0,
\end{align}

\noindent where $E_\text{in}$ ($G_{\text{C, in}})$ is Young's modulus (toughness) of the inclusion material, with $E_\text{0}$ ($G_{\text{C,0}}$) being that of the matrix material. In this work, we limit our attention to inclusions made with the same $\nu$ and $\rho$ as the base medium, but from more compliant materials ($\alpha < 1$) that can also be tougher ($\beta > 1$). We choose glass (a typical brittle material) to be our base material, with $\rho = 2520$ kg/m$^3$, $\nu = 0.25$, as well as $E_0 = 80$ GPa and $G_\text{C,0} = 7.0312$ kg/s$^2$  from \cite{serbena2015crystallization}.

As a point of departure, we consider dynamic fracture under a quasi-static loading condition in this work, so that wave propagation and intertial effects associated with the externally-imposed load may be neglected.  However, this does not imply that the material response is slow once fracture begins\footnote{In contrast, dynamic loadings will induce inertia effects in the form of propagating mechanical disturbances that strike a crack and cause fracture propagation. Spall is a typical example in this regard.}. Brittle fracture can be fast and can cause considerable material inertia effects \cite{achenbach1973quasistatic} especially in the presence of material heterogeneities even if the remote loading is quasi-static \cite{lazzaroni2012role, albertini2021effective}. As such, we use a dynamic phase-field formulation to explicitly account for crack speed and any inertia effects associated with a fast-propagating crack tip. To achieve a quasi-static loading condition while maintaining computational feasibility, we use a loading velocity $v_\text{load} ~ \sim 5 \times 10^{-5}v_R$ for all simulations, where $v_R$ is the Rayleigh wave speed of the glass. (It is generally recommended $v_\text{load} < 1\% v_R$ \cite{smith2009abaqus} to ensure a quasi-static loading condition.)

The last quantity to be determined is $\ell$, a regularization parameter that imposes a length scale on the material heterogeneity; effectively, the smeared crack field is insensitive to material structure at smaller length scales than $\ell$~\cite{hossain2014effective}. As such, we ensure that $d$ and $h$ are strictly larger than $\ell$ for the purpose of this work. Although $\ell$ does not have a strong micromechanical basis, $\ell$ has been shown to set the threshold for crack initiation \cite{bourdin2014morphogenesis, tanne2018crack} and controls the size of the process zone from a cohesive zone interpretation \cite{tran2022cohesive}. As such, it can be argued that $\ell$ serves as a material property parameter governing the crack nucleation strength $\sigma_c$ \cite{tanne2018crack}, which admits the following form under plane stress conditions:

\begin{align}
\sigma_c \propto \sqrt{\frac{EG_\text{C}}{\ell}}.
\label{criticalstress}
\end{align}

As $\ell \rightarrow 0$, the crack nucleation strength goes to infinity, which is consistent with LEFM theory and $\Gamma$-convergence arguments \cite{ambrosio1990approximation}. We tried using the flexural strength of the glass reported in \cite{serbena2015crystallization} ($\simeq 100$ MPa) as $\sigma_c$, but it leads to a very small $\ell \simeq 0.006a$ that is computationally costly for the purpose of this work. This is because a decrease in $\ell$ leads to a decrease in element size (denoted as $\delta$) and subsequently the time step required for a stable dynamic PF simulation.  As such, instead, we pick $\ell =\ell_0 \simeq 0.03a$ as a starting point, and we later decrease $\ell$ (while still maintaining computational feasibility) to check its influence on crack propagation. Note that the rapidly increasing computational demand with decreasing $\ell$ prevents us from probing scenarios with $\ell/a \sim \mathcal{O}(10^{-3})$. Although the simulated material ends up being more ``ductile" than glass, the resulting comparison of crack propagation across different inclusion configurations is valid in a relative sense. Another issue that needs attention is the selection of $\ell$ for the inclusion material, as it will also influence crack nucleation according to Eqn. \ref{criticalstress}. For the majority part of this work, we use more compliant and tougher inclusions (specifically $\alpha = 0.4$, $\beta  = 2.4$) with the same $\ell$ as the base material. This implies that the inclusion material shares (almost) the same $\sigma_c$ with the base material (since $\sqrt{\alpha\beta} \simeq 1$). For a small part of this work, we study complementary cases where inclusions are only more compliant (specifically $\alpha = 0.5, \beta = 1$). We keep $\ell$ uniform, which implies a lower $\sigma_c$ for more compliant inclusions (i.e. easier for crack nucleation), in comparison to that of the base material.

Lastly, we note that imposing $\ell$ through a finite-element discretization leads to a numerical amplification of the fracture toughness, denoted as $G_\text{C}^\text{num}$ \cite{bourdin2008variational}:

\begin{align}
G_\text{C}^\text{num} = G_\text{C}(1+\frac{\delta}{4c_w\ell}),
\label{gnumerical}
\end{align}

\noindent where $c_w$ in the normalization constant presented in Eqn. \ref{pfdensity}, and $\delta$ is the element size. As such, we use $G_\text{C}^\text{num}$ when applying fracture mechanics theories to interpret the simulation results. For a given $\ell$ value, we conduct simulations across different inclusion configurations for the same amount of time $t$ over irregular linear triangular elements. Regardless of the choice of $\ell$, we select elements within the crack propagation region to have a size of $\delta \simeq \ell/3$, which is small enough to resolve the crack evolution with sufficient accuracy \cite{miehe2010thermodynamically}.

\subsection{Representative result on inclusion-modulated crack propagation}\label{repexample}

\begin{figure}[h]
\centering
\includegraphics[width=\linewidth]{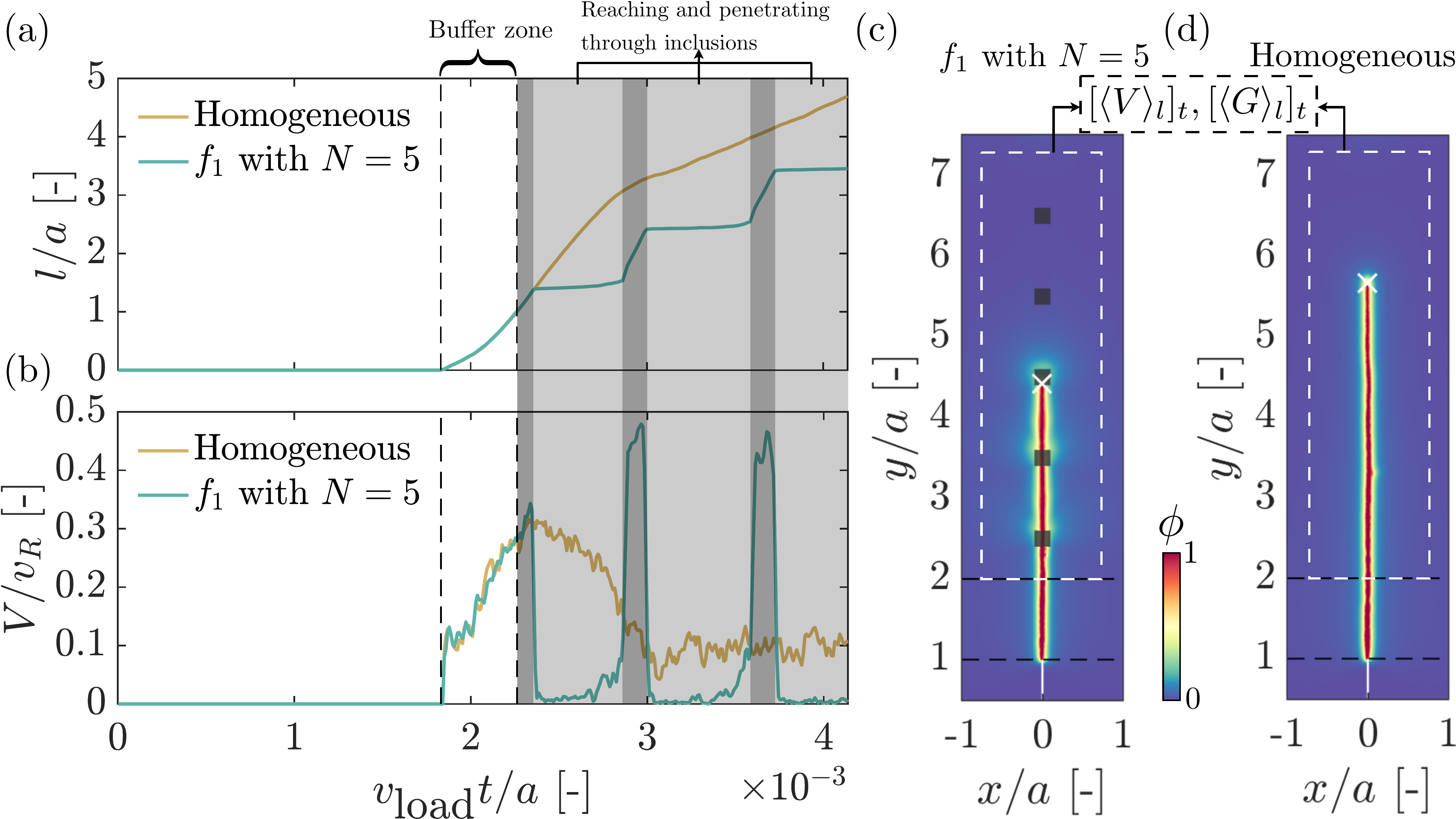}
\caption{(a) Evolution of normalized crack length $l/a$ as a function of normalized indentation displacement $v_\text{load}t/a$ for a homogeneous case with no inclusion (yellow curve) and a heterogeneous case with five inclusions (blue curve, $d = 0.2a$ resulting from setting $c_0 = 0.2, N_0 = 5$). The inclusion material has an elastic contrast $\alpha = 0.4$ and a toughness contrast $\beta = 2.4$. The two alternating shaded areas indicate two different types of crack tip locations for the heterogeneous case. The dark gray area indicates that the crack tip is outside of an inclusion, whereas the light gray area indicates that the crack tip reaches the boundary of or is penetrating through an inclusion.  The (almost) constant crack length within each light gray area indicates that the crack is almost arrested as reaching the boundary of and penetrating through an inclusion. (b) A plot similar to (a) but showing the variation of the normalized instantaneous crack tip speed $V/v_R$, where $v_R$ is the Rayleigh wave speed of the base material. (c) A visualization of the final crack trajectory based on $\phi$ showing the interaction between the crack and the inclusions for the heterogeneous case. The white cross indicates the identified crack tip, and the white dashed rectangle indicates the region within which $[\langle V\rangle_l]_t$ and $[\langle G \rangle_l]_t$ are calculated. (d) A similar visualization as (c) but showing the result of the homogeneous case.}
\label{model}
\end{figure}

As a representative result, Fig. \ref{model}(a) shows the evolution of the normalized crack length ($l/a$) as a function of the normalized indentation displacement ($v_\text{load}t/a$), where we calculate $l$ by tracking the crack tip based on the phase field $\phi$ (locating the tip of an isocurve with a threshold value $\phi= 0.85$\footnote{We have tried varying this threshold value ranging from 0.75 to 0.95, from which the resulting crack tip locations show negligible differences.}, see Appendix for details). Two configurations are included in this figure: a homogeneous one with no inclusion (yellow curve), and a heterogeneous one with five inclusions (green curve, material properties are $\alpha = 0.4, \beta = 2.4$, with $d = 0.2a$ resulting from $c_0 = 0.2, N_0 = 5$). Both simulations are conducted using $\ell = \ell_0$. We observe that cracks initiate at the same time and the evolution of crack lengths remains identical until leaving the buffer zone. This observation is further confirmed by looking at the corresponding (normalized) instantaneous crack tip speed ($V/v_R$, with $V$ calculated using $l$ and $t$), as shown in Fig. \ref{model}(b). Here, $v_R$ is the Rayleigh wave speed of the base material. We confirm that this observation is consistent across all simulated configurations in this work, making it straightforward to ensure that our subsequent analysis isolates the effect of material heterogeneity on dynamic crack propagation. 

After a crack leaves the buffer zone, it enters the region where the inclusions reside. The shaded area in Figs. \ref{model}(a) and (b) with two alternating colors indicate two different regions for crack propagation in the heterogeneous configuration: light grey indicates the region where the crack reaches propagates inside inclusions, while dark gray indicates otherwise. Inclusion-modulated crack propagation is evident as the crack gets (almost) arrested when reaching and penetrating through inclusions: $l$ is (almost) constant and $V$ is (almost) zero. Figs. \ref{model}(c) and \ref{model}(d) visualize the final crack patterns for these two configurations, where the white cross marker in each image indicates the location of the identified crack tip. This observation of the crack propagating through tougher and more compliant inclusions with decreasing speed is consistent with experimental observations reported in a recent work \cite{albertini2021effective}.

\begin{figure}[h]
\centering
\includegraphics[width=\linewidth]{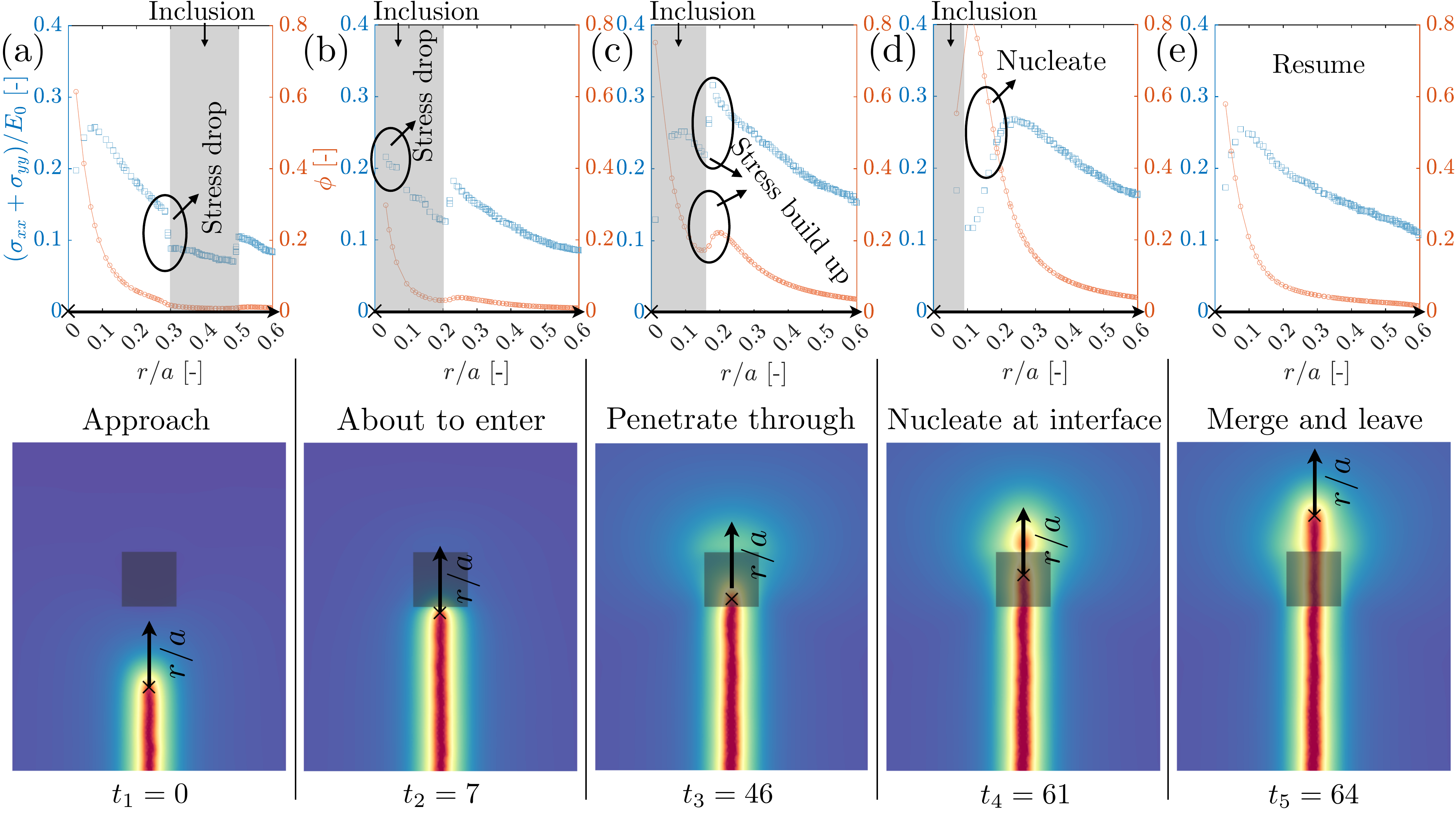}
\caption{Variations of the normalized stress $(\sigma_{xx}+\sigma_{yy})/E_0$ (blue squares) and $\phi$ (red circles) as a function of $r/a$ as the crack approaches (a), about to enter (b), penetrates through (c), nucleates at the interface (d), and merges and leaves (e) an inclusion of size $d = 0.2a$. The grey shaded area in each sub-figure shows the portion of that inclusion ahead of the crack tip (the black cross shown in all sub-figures). The time is set to zero at (a), and the value is relative, used to indicate the rate of crack propagation at each stage. In this particular case, crack entering the inclusion ($t_3-t_2$) takes more time than that for crack nucleating at the compliant-to-stiff interface ($t_4-t_3$) after entering the inclusion. }
\label{stress_evolv}
\end{figure}

 The observed crack-inclusion interaction in the presence of more compliant and tougher inclusions is well documented \cite{hossain2014effective, albertini2021effective}, which can be attributed to the presence of stress fluctuation (for delaying crack entering an inclusion \cite{hsueh2018stress}), the presence of higher toughness and lower modulus (for slowing down crack advancing inside that inclusion \cite{albertini2021effective}), and the presence of crack nucleation when crossing a compliant-to-stiff interface (for delaying the crack leaving that inclusion \cite{zak1962crack, hsueh2018stress}). As an illustrative example, Fig. \ref{stress_evolv} shows the normalized stress distribution, $(\sigma_{xx}+\sigma_{yy})/E_0$ and $\phi$ ahead of the crack tip as it approaches, enters, penetrates through and leaves the first inclusion shown in Fig. \ref{model}(c). On the one hand, the deceleration of the crack as it approaches the inclusion results from decreased stress ahead of the crack tip, causing delayed penetration into the inclusion, as shown in Figs. \ref{stress_evolv}(a) and (b); on the other hand, the stress at the compliant-to-stiff interface builds up as the crack slowly penetrates through the tougher inclusion, as shown in Fig. \ref{stress_evolv}(c). When stress builds up to the nucleation point, a crack also nucleates on the stiffer side of the compliant-to-stiff interface, as shown in Fig. \ref{stress_evolv} (d); then, the crack quickly accelerates and leaves the inclusion by merging with the crack tip inside the inclusion\footnote{Note that whether and when this kind of nucleation and merging happens is affected by the particular choice of material. In general, a larger $\alpha$ promotes such behavior. A simple example is setting inclusion to be void ($\alpha = 0$), for which a crack must renucleate on the other side of the interface.}, as shown in Fig. \ref{stress_evolv}(e). After the crack leaves the crack, the stress resumes its distribution shown in Fig. \ref{stress_evolv}(a) (excluding the part inside the inclusion). 

\subsection{The relevance of the size of the K-dominant zone}\label{kzonerole}

The specific crack-inclusion interaction certainly depends on the particular choice of material. However, leaving the choice of material unchanged, our observation raises a possible effect of the size of the inclusion in terms of altering crack propagation. This kind of size effect certainly exists and should operate on a scale larger than that of the process zone, as already evidenced by the same recent work \cite{albertini2021effective}, in which the crack speed fluctuation and the apparent fracture energy were found to depend on the heterogeneity size. In that work, the length scale of heterogeneities probed is on the order of several millimeters, much larger than that of the process zone of the PMMA-like material used\footnote{This value is estimated based on \cite{irwin1997plastic, susmel2008theory}, using the material properties of that PMMA-like material, i.e. VeroClear. It also agrees with the typical process zone size of glassy polymers ($ < 0.1$ mm) as documented in the literature \cite{LEEVERS20013322}.}: $\sim \mathcal{O}(10^{-2})$ mm. Note that a similar type of size effect can be observed from experiments carried out in \cite{brodnik2021fracture}, where the direction of crack propagation in an architected PMMA sheet depends on the spacing between the voids, both of which are on the millimeter scale and are much larger than the process zone size of PMMA. In our case, the inclusion size $d = 0.2a \simeq 6.7\ell_0$, suggesting that the ``PF crack" senses the inclusions and gets arrested sequentially. Can a smaller inclusion still arrest the crack? If so, is there a lower bound for it? For instance, in a semi-infinite domain a crack has trouble crossing a compliant-to-stiff interface due to decreasing crack tip driving force \cite{zak1962crack}, but will this continue to hold as the stiffer domain shrinks in size?

\begin{figure}
\centering
\includegraphics[width=1.0\linewidth]{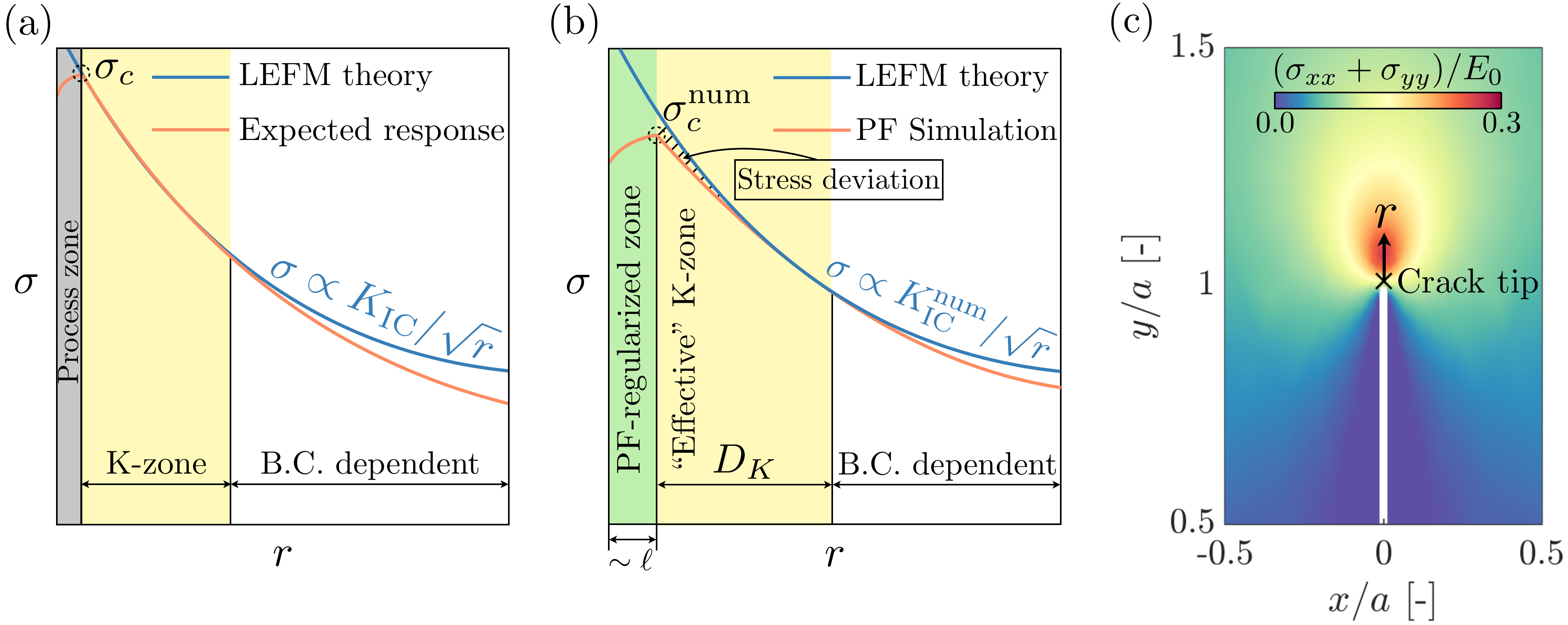}
\caption{(a) A schematic showing the variation of stress $\sigma$ as a function of the distance $r$ to the crack tip, based on either LEFM (blue curve) or the expected response of a brittle material (red curve).  Three regions can be identified. Very close to the crack tip is the process zone (colored in grey) where both dissipation and nonlinear elasticity prevail \cite{chen2017instability}, and LEFM breaks down. Following the process zone is an annulus called the K-dominant zone (colored in yellow) where LEFM holds. At the junction of these two zones, the crack nucleation stress $\sigma_c$ is achieved. After the K-dominant zone is where boundary effects play a role and where LEFM  breaks down again. (b) A schematic resembling (a) but showing the expected outcome from a typical phase-field simulation (red curve). Immediately after the numerically identified crack tip ($r=0$) is the PF-regularized zone (colored in green and with a size of about $\ell$) representing mechanisms active in the process zone \cite{janssen2004fracture}. It ends where the stress peaks at the numerical crack nucleation stress ($\sigma_c^\text{num}$) whose value is related to $\ell$ (see Eqn. \ref{criticalstress}) and $\delta$. Following the PF-regularized zone is the proposed ``effective K-zone" (colored in yellow and with a size denoted as $D_K$) that ends where the PF simulation result deviates from the LEFM solution. The solid line-shaded area after the PF-regularized zone indicates stress deviations resulting from FE discretization errors and non-zero (but small) $\phi$ values. (c) A visualization showing the spatial distribution of $(\sigma_{xx}+\sigma_{yy})/E_0$ at the moment of crack initiation for the homogeneous case obtained using $\ell = \ell_0$. We indicate the crack tip location using the black cross and denote the vertical distance away from the crack tip as $r$.}
\label{kzonediscussion}
\end{figure}

Is there a relevant (or characteristic) length scale associated with this kind of size effect? We hypothesize that it is associated with the K-dominant zone derived from Linear Elastic Fracture Mechanics (LEFM) theory. For brittle materials which satisfy the \textit{small-scale yielding} condition \cite{janssen2004fracture}, a crack propagates by sensing and exploring the stress field inside the K-dominant zone \cite{williams1957stress, brodnik2021fracture}, an annulus that sits in between the much smaller process zone (where both dissipation and non-linear elasticity prevail \cite{chen2017instability}) and an outer zone (where boundary conditions play a role), as shown in Fig. \ref{kzonediscussion}(a). More specifically, although when and where a crack decides to propagate are determined by the \textit{time} and \textit{position} (on the scale of the process zone) at which $\sigma = \sigma_\text{c}$, this specific time and position for $\sigma = \sigma_\text{c}$ depend \textit{non-locally} on a surrounding area (i.e. on the scale of the K-dominant zone where $\sigma \propto K_\text{IC}/\sqrt{r}$ \cite{williams1957stress}).


Consequently, when a crack approaches an inclusion, the size of that inclusion with respect to that of the K-dominant zone becomes relevant. When the inclusion is large enough to encapsulate the K-dominant zone, the crack tip senses a homogeneous material field (corresponding to the inclusion material). When the inclusion size is not large enough to encapsulate the K-dominant zone (but still large enough compared to the size of the process zone), the crack tip senses a heterogeneous material field: part of this field consists of the inclusion and part of this field consists of the base medium. This can lead to a stress modulation within the K-dominant zone, which may change the time and position of which $\sigma = \sigma_\text{c}$ depending on the particular choice of $\alpha$, $\beta$, and inclusion position (relative to the crack path). In our case, once the inclusion size becomes smaller than that of the K-dominant zone, a crack will also sense the base medium (which is stiffer and more brittle) as it approaches and penetrates through an inclusion. This implies that the inclusion may not be able to slow down the crack entering and moving through (due to insufficient stress drop for the former and faster stress build-up for the latter, as the K-dominant zone always contains the base medium),  and it may subsequently accelerate the crack crossing the compliant-to-stiff interface (due to faster stress build up as the K-dominant zone covers that interface). As a result, inclusions with sizes smaller than that presented in Fig. \ref{model} can lead to different crack propagation dynamics.

Therefore, our main interest is to quantify and compare different crack propagation dynamics resulting from the potential interplay between the size of inclusion and that of the K-dominant zone. As shown in Fig. \ref{config}(d), on one end, we can have small but dense inclusions (small $d$ and $h$); on the other end, we can have large but sparse inclusions (large $d$ and $h$). Between $d$ and $h$, which is more important? Should we have fewer large inclusions or more small inclusions? A particularly interesting configuration in the middle can arise: the size of the inclusion meets that of the K-dominant zone with the spacing being determined accordingly. We use two spatio-temporally averaged variables for comparison and quantification: the apparent crack speed $[\langle V\rangle_l]_t$ and the apparent fracture energy dissipation rate $[\langle G\rangle_l]_t$. Here, $\langle \cdot \rangle_l$ denotes the spatial average of a quantity at a given time instant over the crack length $l$, and $[\cdot]_t$ denotes the temporal average of a quantity over every time instant $t$. We do not calculate $V$ and $G$ until a crack leaves the buffer zone. As an example, the two dashed black rectangles shown in Figs. \ref{model}(c) and \ref{model}(d) indicate the regions within which $V$ and $G$ are calculated. As to our calculation procedure. First, at every instant $t$, we can calculate the crack length $l$, and we can also calculate the fracture energy $\Gamma_\ell = \int_\Omega \gamma_\ell \text{d$\Omega$}$. This allows us to compute $\langle V\rangle_l =l/t$ and $\langle G \rangle_l = \Gamma_\ell /l$. Then, we perform an arithmetic average over $t$ to get $[\langle V\rangle_l]_t$ and $[\langle G \rangle_l]_t$. Lastly, for simplicity, we use the following two non-dimensionalized parameters for comparing $[\langle V\rangle_l]_t$ and $[\langle G\rangle_l]_t$ across different configurations:

\begin{align}
\tilde{V} = \frac{[\langle V\rangle_l]_t}{[\langle V_0\rangle_l]_t}, \,\,\text{and}\,\,\tilde{G} = \frac{[\langle G\rangle_l]_t}{[\langle G_0\rangle_l]_t},
\label{normalization}
\end{align}
where $V_0$ and $G_0$ follow the same definition of $V$ and $G$, but they are computed from the homogeneous configuration. Self-evidently, a smaller value of $\tilde{V}$ and a larger value of $\tilde{G}$ indicate a better toughening outcome than that from the homogeneous configuration: lower crack speed and higher fracture energy dissipation rate. 

\section{K-dominant zone in phase-field simulations and the effect of $\ell$}\label{kzonewithl}

\subsection{Theoretical considerations for estimating the K-dominant zone}
Estimating the size of the K-dominant zone is challenging in the presence of (meso-scale) material heterogeneity. Meso-scale material heterogeneities (i.e. inclusions in this work) introduce new boundaries that can be sensed by a crack tip through stress waves, which can further complicate the instant stress state around a crack tip. Here, we use some simplifications to estimate $D_K$ as a first-order approximation. First, we assume negligible changes in the K-dominant zone size for our configurations, and accordingly, we estimate $D_K$ when a crack initiates from the notch tip. Second, we neglect the crack speed effect on the K-dominant solution \cite{freund1998dynamic}, meaning the quasi-static formulation is adopted. Putting together, in this work, we estimate $D_K$ by comparing the stress distribution computed from PF simulations to that from LEFM, in a way similar to \cite{becker1997limitations}.


\subsection{Numerical considerations for defining an ``effective" K-dominant zone} \label{numKzone}
Following the above theoretical considerations, we can overlay the stress distribution from PF simulations onto that from LEFM to estimate the size of the K-dominant zone, i.e., the width of the yellow region in Fig. \ref{kzonediscussion}(a). However, in practice, this leads to numerical challenges that require caution when defining the K-dominant zone. Fig. \ref{kzonediscussion}(b) shows a typical example of stress distribution ahead of a ``PF crack tip", which shows two major differences from the expected stress distribution approaching a sharp crack demonstrated in Fig. \ref{kzonediscussion}(a). First, there is no longer a process zone as our material model is elastic; instead, immediately following the ``PF crack tip" is a PF-regularized zone whose size is about $\ell$. Although this zone does not have a strong micromechanical foundation, $\ell$ is known to control the size of process zone from a cohesive zone perspective, and it can be calibrated according to the physical process zone size \cite{tran2022cohesive}. In parallel, the numerical crack nucleation stress $\sigma_c^\text{num}$ depends on $\ell$ (see Eqn. \ref{criticalstress}) and $\delta$. Note that usually we have $\sigma_c^\text{num} <\sigma_c$ as a result of limited computational capability. Second, after passing $\sigma_c^\text{num}$, the stress can still deviate from the LEFM calculation over a small distance, due to FE discretization errors (in resolving elasticity) and non-zero (although small) $\phi$ values (in degrading material modulus). The detailed pattern of this deviation depends on $\ell/\delta$ and the specific PF model (e.g., AT2 \cite{ambrosio1990approximation} or AT1 \cite{pham2011gradient}), with the latter controlling the spatial variation of $\phi$. In general, decreasing $\ell$ while increasing $\ell/\delta$ better approximates the brittleness of material of interest ($\sigma_c^\text{num} \rightarrow \sigma_c$) and reduces the stress deviation immediately outside of the PF-regularized zone.

Taking both differences into account, we consider an ``effective" K-dominant zone, denoted as $D_K$ herein, which starts from  $\sigma_c^\text{num}$ and ends at the location where the stress deviates from the LEFM prediction, shown as the yellow region in Fig. \ref{kzonediscussion}(c). Note that the LEFM prediction is calculated using $K_\text{IC}^\text{num} = \sqrt{G_\text{C}^\text{num}E}$ with $G_\text{C}^\text{num}$ defined in Eqn. \ref{gnumerical}. We next present several numerical studies aimed at understanding the effect of $\ell$ and $\ell/\delta$ on $D_K$. Since we only need to estimate $D_K$ at the moment of crack initiation, the computational cost is greatly reduced and we can probe into scenarios with $\ell/a \sim \mathcal{O}(10^{-3})$. In practice, we extract the variation of  $(\sigma_{xx}+\sigma_{yy})/E_0$ as a function of $r$ from simulations and compare it with the LEFM prediction $\sigma \propto K_\text{IC}^\text{num}/\sqrt{2\pi r}$. To aid illustration, Fig. \ref{kzonediscussion}(c) shows the spatial distribution of $(\sigma_{xx}+\sigma_{yy})/E_0$ at the moment of crack initiation from a simulation with $\ell = \ell_0$.

\subsection{Varying $\ell$ while holding $\delta$ fixed}

\begin{figure}[H]
\centering
\includegraphics[width=1.0\linewidth]{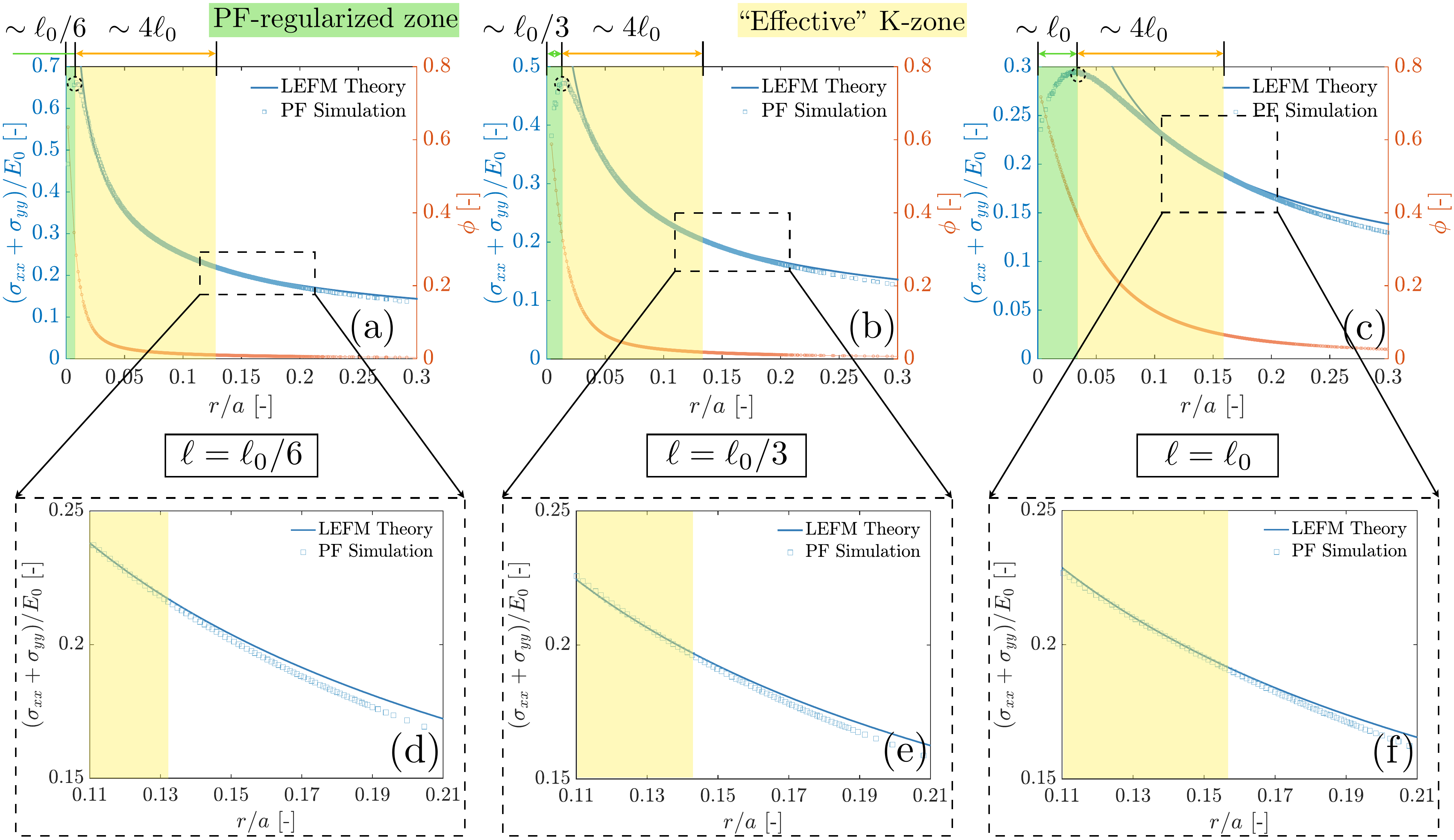}
\caption{Variations of $(\sigma_{xx}+\sigma_{yy})/E_0$ (left axis) and $\phi$ (right axis) as a function of $r$ from the configuration shown in Fig. \ref{kzonediscussion}(c), obtained with fixed $\delta$. Symbols (hollow squares and circles) are from PF simulations, while solid lines are from LEFM calculations.  For all simulations, we fix $\delta \simeq \ell_0/26$ near the notch tip area. The PF-regularized zone and the ``effective K-zone" discussed in Fig. \ref{kzonediscussion}(b) are highlighted using the same color scheme. Results from the left to the right are from different $\ell$ used in simulations: $\ell = \ell_0/6$ for (a) with a zoomed-in plot shown in (d); $\ell = \ell/3$ for (b) with a zoomed-in plot shown in (e), and $\ell = \ell_0$ for (c) with a zoomed-in plot shown in (f).}
\label{kzonecalibration1}
\end{figure}

Figs. \ref{kzonecalibration1}(a), (b), and (c) show the variation of $(\sigma_{xx}+\sigma_{yy})/E_0$ and $\phi$ as a function of $r/a$ using $\ell = \ell_0/6$, $\ell = \ell_0/3$, and $\ell = \ell_0$, respectively. All simulations have fixed $\delta \simeq \ell_0/26$. The point $r/a = 0$ corresponds to the ``PF crack tip", which is defined as the tip of the fitted iso-curve with a value of 0.85 (see Section~\ref{repexample}). These three figures illustrate Fig. \ref{kzonediscussion}(b) in a more quantitative sense. Figs. \ref{kzonecalibration1}(d), (e), and (f) zoom into the region where the stress deviates from the LEFM predictions. The most important observation drawn is that $D_K$, the size of the effective K-zone, does not change appreciably with $\ell$, staying around $D_K \simeq 4\ell_0$. This leads to $D_K \sim \mathcal{O}(10\ell)$ for the first two cases and $D_K \sim \mathcal{O}(\ell)$ for the third case. Decreasing $\ell$ thus seems to simply ``pull" the effective K-zone more closely into the crack tip. Further decreasing $\ell$ is certainly desired for representing $D_K \gg \ell$ encountered in real brittle materials, but it can become computationally intractable. Unlike $D_K$, stress and $\phi$ ahead of the crack tip do change with $\ell$. In general, decreasing $\ell$ increases the peak stress. As expected from any PF model, away from the PF-regularization zone, the value of $\phi$ remains small; as approaching the PF-regularized zone, $\phi$ increases with a gradient that scales inversely with $\ell$: the smaller the $\ell$, the larger the gradient.  The specific variation pattern of $\phi$, on the other hand, depends on the particular PF formulation as mentioned in Section~\ref{numKzone}. In addition, a greater $\phi$ value (resulting from a larger $\ell$) within the effective K-zone leads to a more pronounced stress deviation from the LEFM prediction since the material degradation is higher for larger $\phi$.

\subsection{Varying $\ell$ while holding $\ell/\delta$ fixed}
\begin{figure}[H]
\centering
\includegraphics[width=0.8\linewidth]{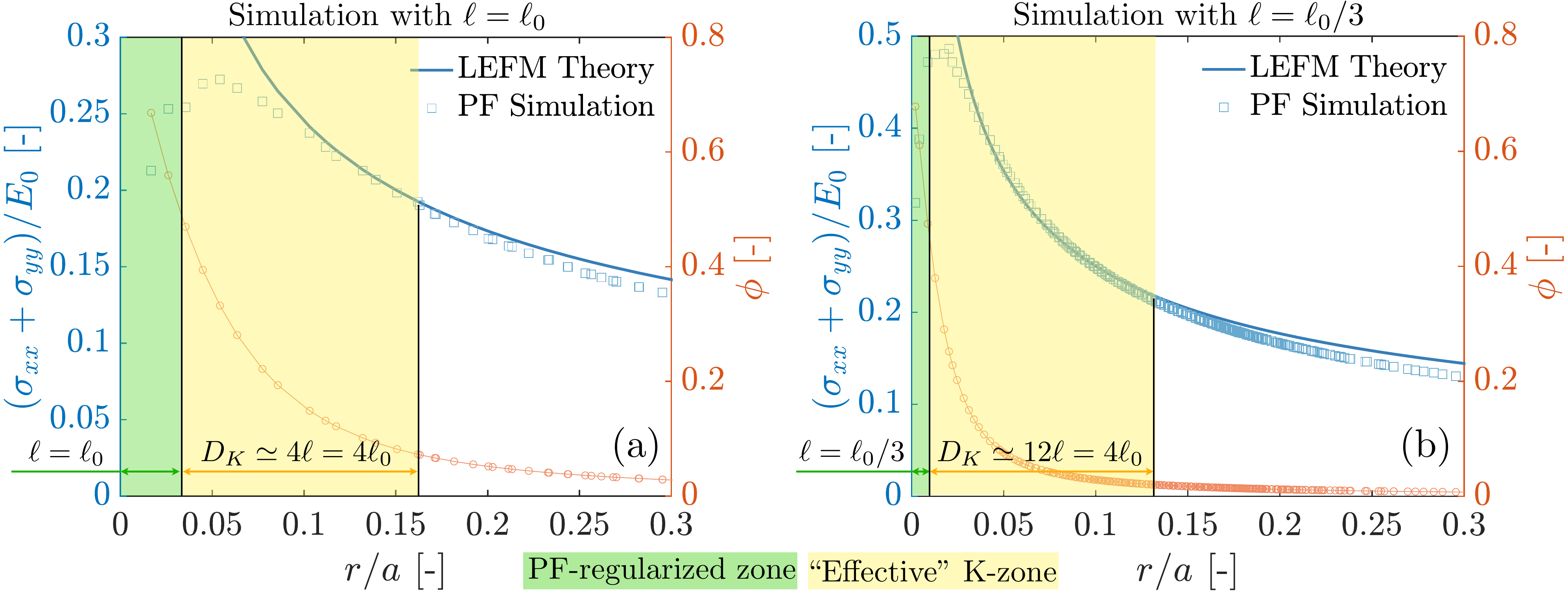}
\caption{Variations of $(\sigma_{xx}+\sigma_{yy})/E_0$ (left axis) and $\phi$ (right axis) as a function of $r$  from the configuration shown in Fig. \ref{kzonediscussion}(c), obtained with fixed $\ell/\delta$. Symbols (hollow squares and circles) are from PF simulations, while solid lines are from LEFM calculations. For all simulations, we fix $\ell/\delta \simeq 3$ near the notch tip area. From left to right are results from different $\ell$ used in simulations: $\ell = \ell_0$ for (a) and $\ell = \ell_0/3$ for (b).}
\label{kzonecalibration2}
\end{figure}

Figs. \ref{kzonecalibration2}(a) and (b) show the variation of $(\sigma_{xx}+\sigma_{yy})/E_0$ and $\phi$ as a function of $r/a$ using $\ell = \ell_0$ and $\ell = \ell_0/3$, respectively. Now both simulations have fixed $\ell/\delta \simeq 3$, implying that $\delta$ used in these two scenarios are larger compared to those used for Fig. \ref{kzonecalibration1}(b) and (c). Again, increasing $\ell$ decreases the peak stress, broadens the PF-regularized zone, and increases the stress deviation and $\phi$ in the effective K-zone; all are consistent with observations made in the previous section. Most importantly, $D_K$ again does not change appreciably, staying around $4\ell_0$ in both cases, thereby suggesting the relative insensitivity of $D_K$ to both $\ell$ and $\delta$. Overall, these observations imply that although changing $\ell$ or $\ell/\delta$ can change the variation of $\tilde{V}$ and $\tilde{G}$, the comparison of these two quantities across different configurations should remain valid in the relative sense. In light of this, we can run full simulations (instead of concerning just crack initiation) using relatively large $\delta$ (while maintaining $\ell/\delta > 2$ \cite{miehe2010thermodynamically}) to save significantly the computational expenses. In practice, we perform two sets of full simulations studying crack-inclusion interactions using $\ell = \ell_0$ with $\ell/\delta \simeq 3$ and $\ell = \ell_0/3$ with $\ell/\delta \simeq 2.2$, respectively. All simulations are run for the same amount of loading time.

\section{Result and discussion}\label{result}

\subsection{Identical area fraction of inclusions with $\alpha = 0.4, \beta = 2.4$} \label{result1} 
We discuss results obtained from configurations with identical area fractions of inclusions. We find $c_0 = 0.2, N_0 = 5$ is a good option to generate configurations that cover the crossover $d = D_K$, where we set $D_K = 4\ell_0$ based on discussions from Section~\ref{kzonewithl}. Using  $c_0 = 0.2, N_0 = 5$, the resulting inclusion size ranges from as small as $\sim 0.06a$ ($2\ell_0$) to as large as $\sim 0.26a$ ($8.7\ell_0$). We fix the choice of inclusion material to be $\alpha = 0.4, \beta = 2.4$.

\begin{figure}[H]
\centering
\includegraphics[width=\linewidth]{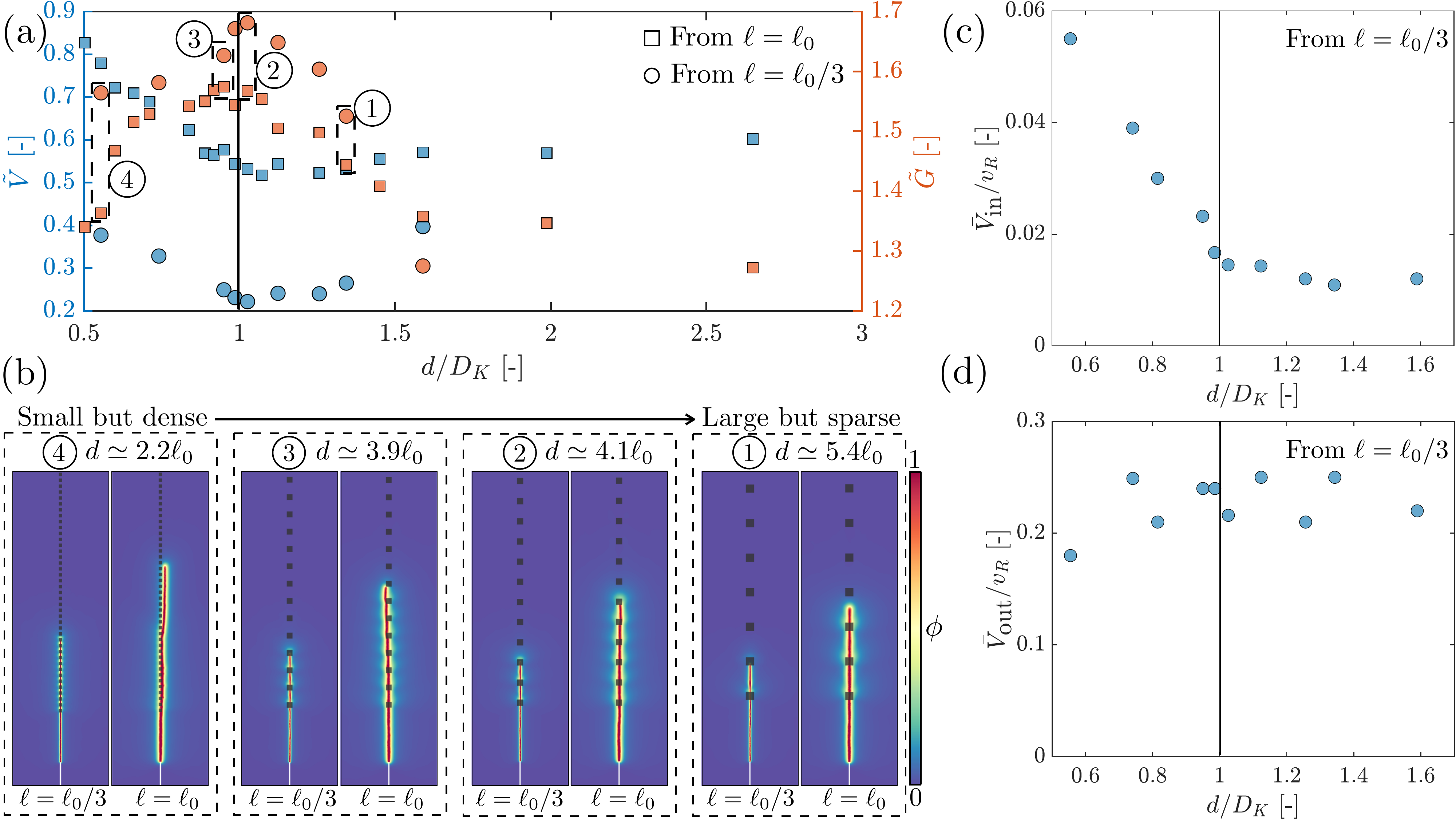}
\caption{Results under identical area fraction $f_2 = $ const. (a) Variations of $\tilde{V}$ (left axis) and $\tilde{G}$ (right axis) as a function of normalized inclusion size $d/D_K$ where $D_K = 4\ell_0$. Data points represented by squares are from simulations with $\ell = \ell_0$ while those represented by circles are from simulations with $\ell = \ell_0/3$. Data points encompassed by the black solid rectangle are from the configuration with $f_1|_{N=5}$. Four configurations are highlighted by black dash rectangles. (b) Snapshots of the crack trajectory, picked at the same loading time, for the four configurations highlighted in (a). (c) Average crack speed $\bar{V}_\text{in}$ inside inclusions. (d)  Average crack speed $\bar{V}_\text{out}$ outside inclusions.}
\label{0424f2result}
\end{figure}

Fig. \ref{0424f2result}(a) shows the variation of $\tilde{V}$ and $\tilde{G}$ as a function of $d/D_K$ for $\ell = \ell_0$ (squares) and $\ell = \ell_0/3$ (circles). The quantitative difference between these two sets of data comes from varying $\ell$, and it is partly due to the crack deflecting away from inclusions whose $d$ becomes close to $\ell$. Fig. \ref{0424f2result}(b) shows that the crack deflects away from inclusions in configuration four ($d \sim 2.2\ell_0$) when using $\ell = \ell_0$; however, such behavior is completely gone when using $\ell = \ell_0/3$. Our model setup is completely symmetric, and we believe the observed crack deflection is therefore a numerical issue associated with $d/\ell \simeq 1$, which could be further amplified by wave propagations in modulating stress and finite-element discretization bias in maintaining a perfectly vertical crack path. For simulations done with $\ell = \ell_0$, the numerical issue becomes pronounced when $d/\ell_0$ goes below 4, but we find that it could also happen for a different $d/\ell_0$ value depending on the particular choice of $\alpha$ and $\beta$. We leave an in-depth investigation in this regard for future work, but in general, this numerical issue should vanish as $\ell \rightarrow 0$.

Nevertheless, both sets of data show similar patterns regardless of $\ell$: varying $d$ monotonically leads to non-monotonic variations of $\tilde{V}$ and $\tilde{G}$. When $d>D_K$ (the green region), decreasing the inclusion size $d$ leads to better toughening outcomes: $\tilde{V}$ (slowly) decreases, and $\tilde{G}$ increases. However, the trend is reversed as $d$ keeps decreasing to be smaller than $D_K$ (the orange region). The best toughening outcome therefore emerges around the configuration with $d  = D_K$ that is neither very small but dense nor very large but sparse. Let us expand the discussion in terms of $\tilde{V}$ which is an average representation of crack speed inside and outside inclusions. Towards the small but dense end ($d$ shrinks away from $D_K$),  $\tilde{V}$ increases with decreasing $d$. This implies that the inclusion is not large enough to effectively arrest the crack, and increasing $d$ overweighs decreasing $h$ in slowing down a crack. On the large but sparse end ($d$ grows away from $D_K$), $\tilde{V}$ (slowly) increases with increasing $d$. This implies that although the inclusion is large enough to effectively arrest the crack (e.g., Fig. \ref{model}(b)), decreasing $h$ overweighs increasing $d$ in slowing down a crack. Right in the middle where $d=D_K$, an optimal balance is achieved between $d$ (for crack arrest) and $h$ (for crack advance) that collectively give rise to the smallest $\tilde{V}$.

Figs. \ref{0424f2result}(c) and (d) provide a clearer interpretation, each of which shows the average crack tip speed (normalized by the Rayleigh wave speed of the base medium) inside ($\bar{V}_{\text{in}}$) and outside ($\bar{V}_{\text{in}}$) of inclusions, respectively. Inside inclusions, $\bar{V}_{\text{in}}$ remains a small value of around 0.01 and increases very slowly with the decrease of $d$ when $d >D_K$. This weak dependence on $d$ can be understood as the increasing portion of base material inside the K-dominant zone for a crack penetrating an inclusion with a smaller size at a given instant. However, it rapidly increases once $d$ decreases into $d < D_K$, implying a strong dependence of crack speed on the portion of the base medium with respect to the inclusion material inside the K-dominant zone. In contrast, outside inclusions, $\bar{V}_\text{out}$ is largely unchanged, averaging to about 0.22. On the very small but dense side (i.e., configuration $\#$4), $\bar{V}_{\text{out}}$ goes below 0.2, which may be due to minor coupling effects between adjacent inclusions. In other words, the crack tip senses more than one inclusions as they are close enough to be within the K-dominant zone at the same time. To this see, we can compute the lower bound of $h$ (or $d$) for no such coupling: $h = L_\text{in}/N > D_K$ (and thus $d > c_0\sqrt{L_\text{in}D_K/N_0}$). A quick calculation gives $d > 2.24\ell_0$. For configuration $\#4$, $d \simeq 2.2 \ell_0 < 2.24\ell_0$. Consequently, the crack tip senses the second inclusion once it enters the first one. Further reducing the inclusion size can lead to a stronger coupling effect as a crack tip may sense more inclusions\footnote{This can be however computationally demanding to study via PF simulations because the regularization length scale $\ell$ needs to remain small compared to $d$.}. Indeed, taking the extreme of merging all inclusions as one single strip ($h \rightarrow 0$), the crack speed, which is now the same for $\bar{V}_{\text{in}}$ and $\bar{V}_{\text{out}}$, should average to be somewhere larger than $\bar{V}_\text{in}$ shown in Figs. \ref{0424f2result}(c) but smaller than $\bar{V}_\text{out}$ shown in Figs. \ref{0424f2result}(d).

Overall, these two figures show the interesting threshold set by $D_K$: as long as $d>D_K$, having more inclusions will be more efficient in slowing down a crack. A similar argument could be made by analyzing the variation of $\tilde{G}$: once $d$ is large enough ($d>D_K$), increasing the number of inclusions (i.e., reducing $d$ and $h$) becomes more important. The size of the K-dominant zone may thus be viewed as the minimum size for an inclusion to interact effectively with a crack. As a result, if given the same area fraction of the second-phase material, it is more efficient, in terms of increasing fracture resistance, to increase the number of inclusions that are available to interact effectively (so long as $d > D_K$) with a crack. We conclude by pointing out that this kind of size effect only becomes pronounced when the inclusion size becomes comparable to that of the K-dominant zone (e.g., $0.5 < d/D_K<3$ in this case). This range sits between the two extremes: zero and infinity. If $d$ is way smaller than $D_K$ (approaching zero), we shall consider these inclusions as sub-mesoscale heterogeneities whose contributions may be treated in a homogenized manner. As such, the crack tip effectively senses a homogeneous material field. If $d$ is way larger than $D_K$ (approaching infinity), we can consider these inclusions as layered heterogeneities, within each layer, the crack always senses a homogeneous material field for the most part (except when approaching the interface within a distance $\sim D_K$). This is because the K-dominant zone is almost always encompassed by $d$. From this perspective, the size effect with regard to the K-dominant zone is reminiscent of that occurs at a smaller length scale \cite{barras2017interplay}: the material heterogeneity becomes comparable to the size of the process zone, subsequently leading to a significant perturbation of the rupture dynamics.

\subsection{Identical area fraction of inclusions without toughness contrast ($\alpha = 0.5, \beta = 1$)}\label{result3}

Lastly, we turn to complementary cases where there is no toughness contrast ($\beta = 1$), to investigate the interplay between $d$ and $D_K$. Studies on layered materials \cite{ming1989crack, hsueh2018stress} showed that a crack enters a stiff-to-compliant interface easily but has trouble exiting a compliant-to-stiff interface. Therefore scenarios with $\alpha > 1$ should be representative of those with $\alpha < 1$ in terms of studying the interplay between $D_K$ and $d$, in that a crack approaching a more compliant inclusion is similar to a crack leaving a stiffer one, and vice versa. In light of this, we choose more compliant inclusions (specifically $\alpha = 0.5$) here mainly for numerical convenience, in that the same time step (for simulations with $\alpha = 0.4, \beta = 2.4$) can be used, which also makes the normalization in Eqn. \ref{normalization} straightforward. Stiffer inclusions ($\alpha > 1$), on the other hand, may require a smaller time step to maintain numerical stability, thereby increasing also the computational expense. We focus on the same design used in Section~\ref{result1}, considering $d$ that ranges from $d \sim 0.06a (2\ell_0)$ to $d \sim 0.26a (8.7\ell_0)$. Finally, $\ell = \ell_0/3$ is used for all simulations. Note that this suggests the crack nucleation is easier inside the inclusions compared to the base medium.

\begin{figure}[H]
\centering
\includegraphics[width=1.0\linewidth]{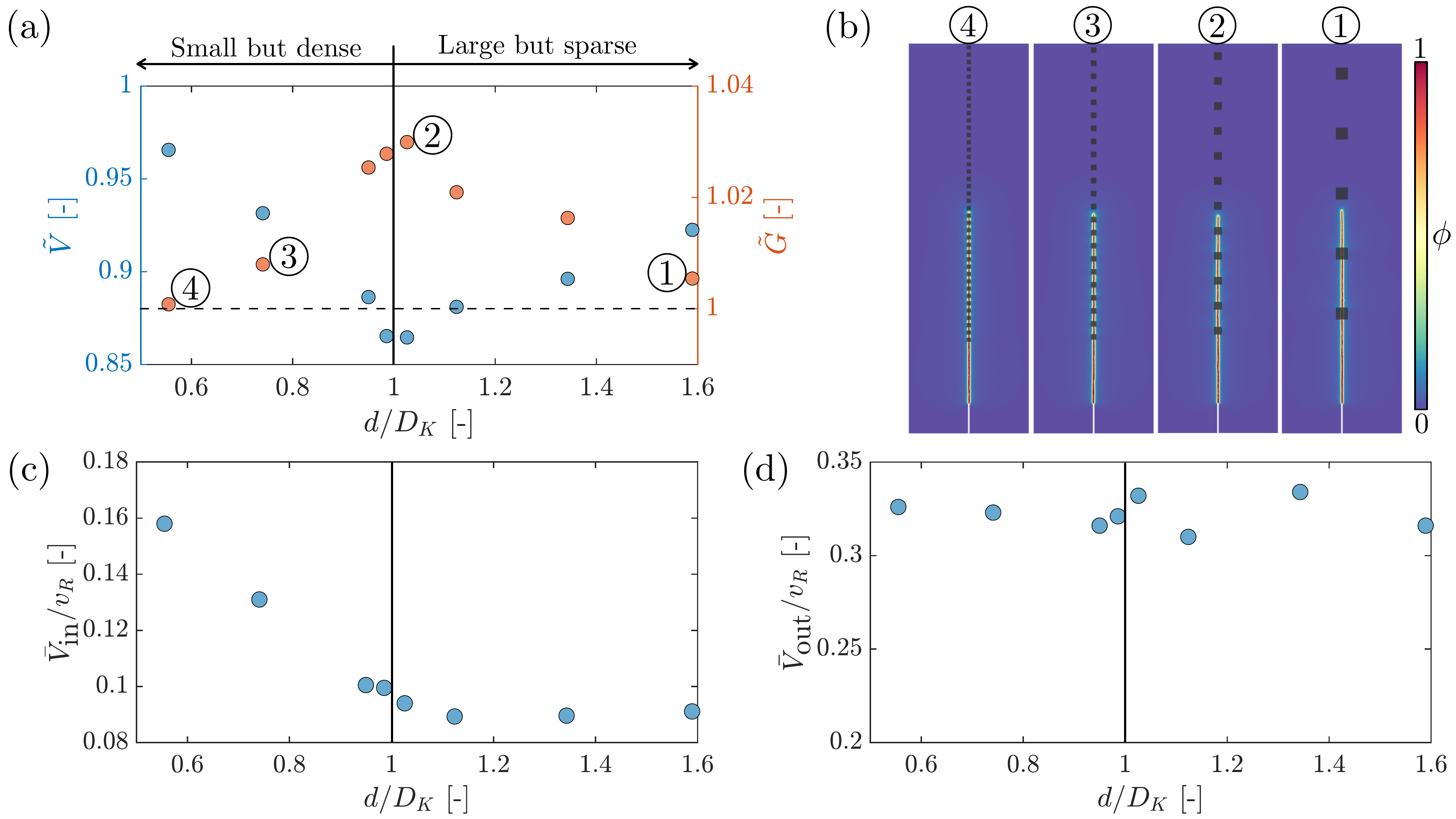}
\caption{(a) Variations of $\tilde{V}$ (left axis) and $\tilde{G}$ (right axis) as a function of normalized inclusion size $d/D_K$. Four configurations are highlighted. (b) Snapshots of the crack trajectory, picked at the same loading time, for the four configurations highlighted in (a). (c) Average crack speed $\bar{V}_\text{in}$ inside inclusions. (d)  Average crack speed $\bar{V}_\text{out}$ outside inclusions.}
\label{0510results}
\end{figure}

For these scenarios, stress fluctuation still exists as long as $\alpha \neq 1$, and therefore $D_K$ should still play a role in modulating crack-inclusion interactions. This is confirmed by comparing Figs. \ref{0510results}(a), (c), and (d) with Figs. \ref{0424f2result}(a), (c), and (d): the best toughening outcome occurs at $d/D_K = 1$, and the average crack speed inside inclusions is largely independent of $d$ for $d >D_K$, but it rapidly increases as soon as $d$ decreases to be smaller than $D_K$. Since inclusions here are no longer as tough as those in Section~\ref{result1} ($\beta = 2.4$), they are not as effective in terms of delaying crack propagation, and therefore they collectively give rise to larger $\tilde{V}$ and smaller $\tilde{G}$ (see Fig. \ref{0424f2result}(a) for a comparison). Another relevant factor is that a crack is easier to nucleate inside the inclusions as we set $\ell$ to be uniform. Decreasing $\ell$ for the inclusion material will delay crack nucleation, and we expect it to shift $\tilde{V}$ downward and $\tilde{G}$ upward\footnote{Again, this becomes computationally challenging for us to verify due to changes needed in decreasing $\delta$ and simulation time step at the same time.}. Fig. \ref{0510results}(b) visualizes the final crack trajectory for four configurations highlighted in Fig. \ref{0510results}(a). Note that for configuration $\#4$, $\tilde{G}$ is very close to one. This may be attributed to coupling effects induced by multiple inclusions enclosed by the K-dominant zone, which we have discussed in Section~\ref{result1}. More specifically, the crack tip has trouble interacting with inclusions one at a time, and in particular, it can sense simultaneously the compliant-to-stiff interface (which slows down crack propagation) and the adjacent stiff-to-compliant interface (which does not slow down crack propagation), leading to less effective toughening.

\section{Summary}\label{sum}
Using a variational phase-field approach, we investigate mesoscale size effects of material heterogeneities on dynamic crack propagation in brittle solids under quasi-static loading conditions. We consider a simple case using a single array of square inclusions to represent mesoscale heterogeneities. We study how altering their geometrical and mechanical configurations can lead to different crack propagation dynamics under a Mode-I loading condition. We summarize our main findings below:

\begin{itemize}

\item When $\ell \simeq d$, numerical issues can arise in that a crack deflects away from inclusions even though our problem setup is entirely symmetric. Finite-element discretization likely plays a role as well: a slightly skewed crack path due to element bias can be amplified by elastic mismatch (as well as wave propagation) when approaching an inclusion. Therefore, caution must be taken in applying PF models to study heterogeneous materials; when possible, it is desired to ensure that the scale of heterogeneity is much larger than $\ell$.

\item Changing $\ell$ does not change appreciably the size of the effective K-dominant zone ($D_K$), but it does change the size of the PF-regularized zone.

\item Results from using $\ell_0$ and $\ell_0/3$ both suggest an interesting size interplay between the size of the inclusion and that of the K-dominant zone in the presence of elastic and toughness contrast. Fixing the area fraction of inclusions and matching the inclusion size $d$ with the size of the K-dominant zone ($D_K$) appears to give the best toughening outcome. This size interplay between $d$ and $D_K$ is still operative when there is only elastic contrast.

\item For scenarios where elastic contrast is present, the size of the K-dominant zone may be viewed as the minimum size for an inclusion to interact effectively with a crack. Therefore, to toughen a material in these scenarios, it is more efficient to increase the number of inclusions available to interact effectively with a crack (i.e., so long as $d \geq D_K$).

\end{itemize}

One limitation of our work lies in the length scale $\ell$ used in PF models. Although $\ell$ does not seem to change qualitatively the conclusions of our study (e.g., size interplay between $d$ and $D_K$), it will change quantitatively the toughening outcome (i.e., $\tilde{V}$ and $\tilde{G}$). The numerical issue tied to $\ell \simeq d$ should vanish as $\ell \rightarrow 0$, but we are unable to verify it at this stage due to excessive computational expenses. Alternatively, how small should $\ell$ be compared to $d$ for simulation results to be representative of real materials? A convergence analysis considering a crack interacting with a single inclusion can be computationally tractable and thus help answer this question. Still, the length scale set by $\ell$ and the associated excessive computational cost make it very challenging to probe into configurations with very small $d$. As such, performing fracture experiments using 3D-printed samples with well-controlled heterogeneity geometry (in a way similar to \cite{albertini2021effective}) should be helpful for testing the general applicability of our work to real brittle materials. In particular, such experiments will be useful for studying different inclusion geometries. An inclusion that is not square-shaped will likely change the stress distribution within the K-dominant zone, leading to different crack-inclusion interactions and toughening outcomes.

It will also be interesting to study scenarios with only toughness contrast. In these scenarios, the stress fluctuation caused by elastic mismatch no longer exists, and the stress distribution inside the K-dominant zone should not depend on the size of the inclusion. So, as a crack approaches and penetrates through an inclusion, the stress distribution ahead of the tip will not change and the response (e.g., the crack speed inside inclusions) should be independent of inclusion size. The interplay between $d$ and $D_K$ may only become relevant in terms of influencing crack nucleation at the interface by modulating how fast the stress builds up to the crack nucleation stress (which can differ between the inclusion and base material).  It will be interesting to see what does the average response looks like and the effect of $\ell$ in this regard. Additionally, in reality, when the inclusion becomes large enough a weak interface can emerge between the inclusion and the base medium, and it can trap an incoming crack if its toughness is small enough \cite{ming1989crack}.   Incorporating a cohesive zone model \cite{rezaei2021direction} will be useful in accounting for these interfacial effects that can become pronounced for curved interfaces \cite{aranda2023crack}, especially in a three-dimensional setting \cite{lebihain2020effective}.

Lastly, it will be helpful to develop a more systematic and efficient approach to estimate $D_K$: given an estimation of the application's and (mesoscale) defects' geometry and size, how do we quickly give a reliable estimation on the range of $D_K$ considering crack evolution? Further, how significant is the effect of fluctuating crack speed on $D_K$?  Previous work has shown that increasing crack speed leads to decreasing the size of the process zone \cite{barras2017interplay}, which implies a possible change to $D_K$ as well. Does such a change of $D_K$ require a modification of the optimal design strategy obtained from a quasi-static estimation? In our work, the crack speed fluctuation is not significant due to a quasi-static loading condition (partly due to our choices of inclusion materials). Therefore, estimating $D_K$ using a quasi-static approximation seems accurate enough from a design perspective. However, the crack speed can fluctuate significantly for dynamic loading conditions and even approach the Rayleigh wave speed when interacting with heterogeneities. In these scenarios, a quasi-static approximation may not hold, and a dynamic formula considering the effect of instantaneous crack tip speed is needed \cite{freund1998dynamic, ravi1982experimental}. Moreover, early experiments have suggested the possible lack of the K-dominant zone due to the highly transient nature of the crack tip motion \cite{ravi1982experimental, tippur1991optical} under dynamic loadings, whose implication toward our finding merits further investigation. Studying the variation of $D_K$ as a function of different loading rates (as well as crack speed) can therefore be helpful for applications in extreme conditions such as high-speed impact.

\section{CRediT Authorship Contribution Statement}
\textbf{L. Li}. Conceptualization, Methodology, Software, Numerical Simulation, Formal Analysis, Investigation, Visualization, Writing -- Original \& Draft. \textbf{J. Rao}. Methodology -- Crack tip tracking algorithm, Writing -- Appendix B. \textbf{TC Hufnagel}. Resources, Supervision, Funding acquisition, Project Administration, Writing -- Review \& Editing. \textbf{KT Ramesh}. Conceptualization, Resources, Supervision, Funding Acquisition, Project Administration, Writing -- Review \& Editing.

\section{Acknowledgements}
Computational resources for this work were provided by the Advanced Research Computing at Hopkins (ARCH) core facility  (rockfish.jhu.edu), which is supported by the National Science Foundation (NSF) grant number OAC1920103. The authors gratefully acknowledge the financial support provided by the Corning Research and Development Corporation, and for stimulating discussions on glass ceramics with Dr. Jason Harris, Dr. Charlene Smith, and Dr. Xinyi Xu from Corning

\appendix

\section{Implementation and verification of our phase-field simulator}
Following \cite{miehe2010phase} for modeling fracture in brittle solids, we decompose the elastic strain energy shown in Eqn. \ref{pfintegral} to a tensile part ($``+$") and a compressive part ($``-"$), with the phase-field acting only on the former:
\begin{align}
\mathcal{W}^e(\epsilon_{ij},\phi) = [(1-k)(1-\phi)^2+k]\mathcal{W}^{e,+}(\epsilon_{ij})+\mathcal{W}^{e,-}(\epsilon_{ij}),
\label{energydecompose}
\end{align}
where $\epsilon_{ij} = (u_{i,j}+u_{j,i})/2$ is the infinitesimal strain tensor, and $k$ is a user-defined small constant used for numerical convenience, preventing $\mathcal{W}^{e,+}$ from vanishing as $\phi \rightarrow 1$. To compute $\mathcal{W}^{e,+}$ and $\mathcal{W}^{e,-}$, we first calculate the tensile part $\epsilon^{+}$ and the compressive part $\epsilon^{-}$ of $\epsilon$ using spectral decomposition \cite{miehe2010phase}:
\begin{align}
\epsilon_{ij} &= \epsilon^{+}_{ij}+ \epsilon^{-}_{ij},\\
\text{with}\quad\epsilon^{+}_{ij} &= \sum_{d=1}^{2} \langle \epsilon^{d} \rangle_+ n^{d}_i n^{d}_j,\\
\text{and}\quad\epsilon^{-}_{ij} &= \sum_{d=1}^{2} \langle \epsilon^{d} \rangle_- n^{d}_i  n^{d}_j,
\end{align}
where $\epsilon^d$ is the $d$-th eigenvalue of $\epsilon$,  $n^d$ is the corresponding eigenvector, $\langle x \rangle_+$ stands for $(x +|x|)/2$, and  $\langle x \rangle_-$ stands for $(x -|x|)/2$ with $|x|$ being the absolute value of $x$. We can then express $\mathcal{W}^{e,+}(\epsilon_{ij})$ and $\mathcal{W}^{e,-}(\epsilon_{ij})$ as the following:
\begin{align}
\mathcal{W}^{e,+}(\epsilon_{ij}) &= \frac{1}{2}\lambda \langle \epsilon_{kk} \rangle_{+}^2+\mu\epsilon^{+}_{kj}\epsilon^{+}_{jl}\delta_{kl}, \label{energyplus}\\
\mathcal{W}^{e,-}(\epsilon_{ij}) &= \frac{1}{2}\lambda \langle \epsilon_{kk} \rangle_{-}^2+\mu\epsilon^{-}_{kj}\epsilon^{-}_{jl}\delta_{kl}, \label{energyminus}
\end{align}
where $\lambda$ and $\mu$ are the Lam\'e constants that can be determined from the Young's modulus $E$ and the Poisson's ratio $\nu$. Applying the principle of least action to Eqn. \ref{pfintegral} with $\mathcal{W}^e$ expressed using Eqns. \ref{energydecompose}, \ref{energyplus} and \ref{energyplus}, we arrive at the following two governing equations:
\begin{align}
\sigma_{ij,j}+b_i &= \rho \ddot{u}_i,\\
\left[1+\frac{4c_w\ell(1-k)}{G_\text{C}}\mathcal{W}^{e,+}\right]\phi-\ell^2\phi_{,ii} &= \frac{4c_w\ell(1-k)}{G_\text{C}}\mathcal{W}^{e,+},
\label{governing}
\end{align}
where $\sigma_{ij} = \partial \mathcal{W}^{e}/\partial {\epsilon_{ij}}$. We enforce the irreversible growth condition $\dot{\phi} > 0$ using a strain-history field \cite{miehe2010phase} over the simulation domain:
\begin{align}
\mathcal{H}(x,t) = \max_{s \in [0,t]}\mathcal{W}^{e,+}\left(  \epsilon(x,s)  \right)\,\,\forall\, x \in \Omega.
\label{historyfield}
\end{align}
Replacing $\mathcal{W}^{e,+}$ with $\mathcal{H}(x,t)$ in Eqn. \ref{governing} we then want to solve:
\begin{align}
\sigma_{ij,j}+b_i &= \rho \ddot{u}_i, \label{ufield}\\
\left[1+\frac{4c_w\ell(1-k)}{G_\text{C}}\mathcal{H} \right]\phi-\ell^2\phi_{,ii} &= \frac{4c_w\ell(1-k)}{G_\text{C}}\mathcal{H},
\label{pdephi}
\end{align}
together with the following Neumann boundary conditions (plus any existing Dirichlet boundary conditions) :
\begin{align}
\sigma_{ij}n_j &= t_i\,\,\text{on}\,\, \partial \Omega,\\
\phi_{,i}n_i &= 0\,\,\text{on}\,\, \partial \Omega.
\end{align}

\begin{figure}[h]
\centering
\includegraphics[width=1.0\linewidth]{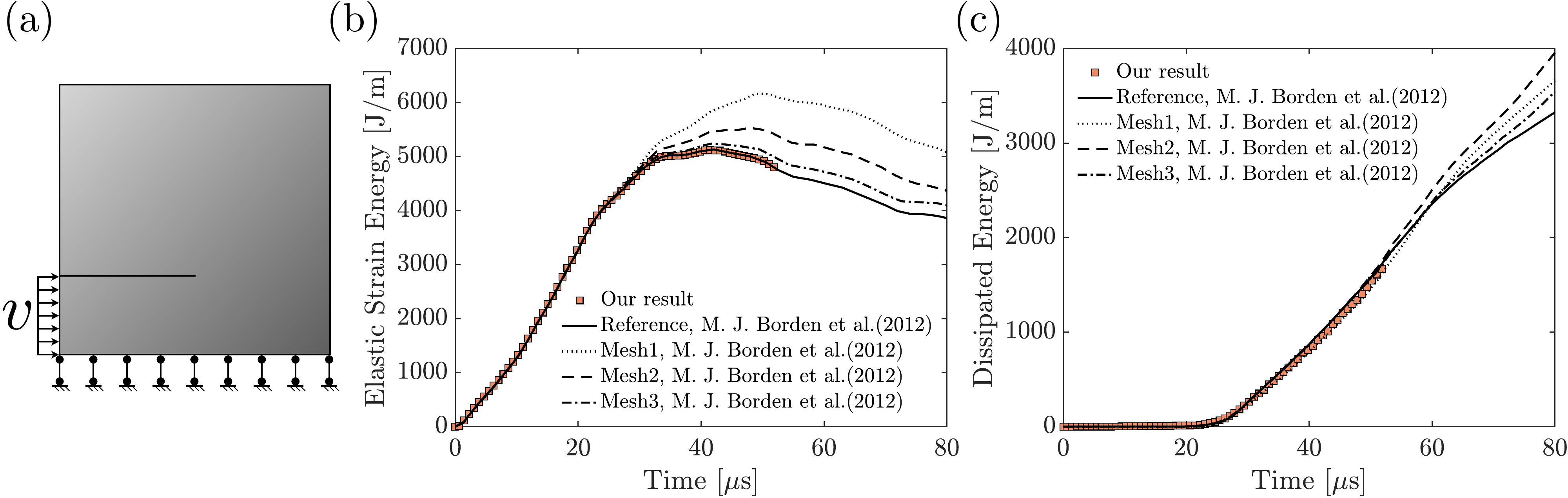}
\caption{(a) Geometry and boundary condition of the simulation domain. (b) Temporal evolution of the system's kinetic energy obtained from our simulator (red circle) and from \cite{borden2012phase} using various element sizes (curves colored in black). (c) A similar plot to (b) but showing the temporal evolution of the system's dissipated energy through fracture.}
\label{kalthoffdata}
\end{figure}

\begin{figure}[h]
\centering
\includegraphics[width=1.0\linewidth]{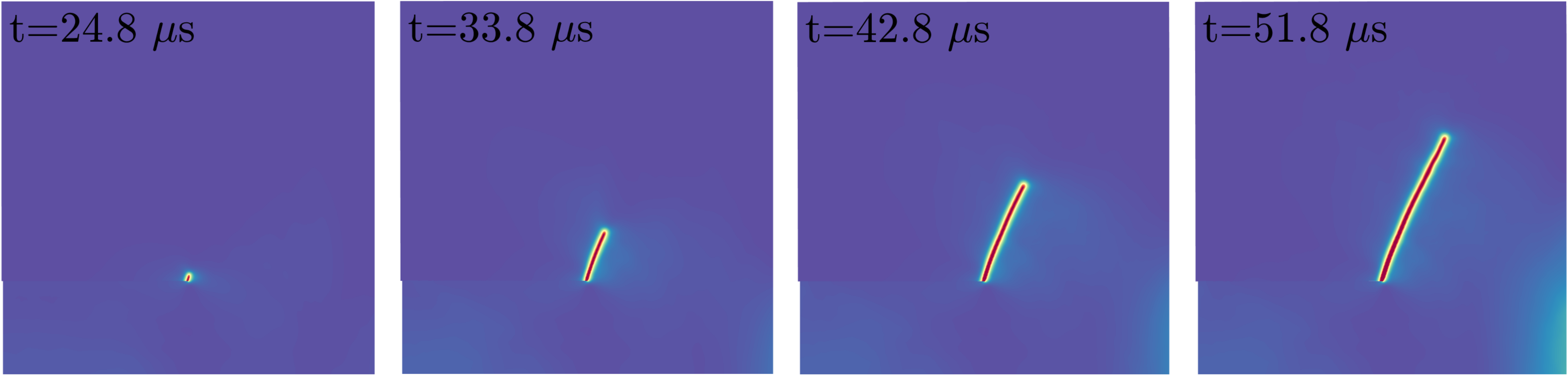}
\caption{Snapshots obtained from our simulation showing the crack trajectory at four different time instants: $t = 24.8$ $\mu$s,  $t = 33.8$ $\mu$s,  $t = 42.8$ $\mu$s, and  $t = 51.8$ $\mu$s.}
\label{kalthoffgraph}
\end{figure}

We note that Eqn. \ref{pdephi} is in fact the same as Eqn. 9 presented in \cite{borden2012phase}. This can be easily checked by redefining $\phi = 0$ as fully damaged and $\phi = 1$ as fully intact and by inserting $c_w = 1/2$ and substituting $\ell = 2\epsilon$. We solve Eqns. \ref{ufield} and \ref{pdephi} weakly following a standard finite element discretization and calculation procedure, using the alternating minimization (or staggered) scheme. We verify our implementation using published simulation results of the classical Kalthoff-Winkler experiment \cite{kalthoff2000modes}. Fig. \ref{kalthoffdata}(a) shows the simulation domain and boundary condition where we also take advantage of the symmetric nature of the experiment to reduce computational cost. We model the impactor by applying the following velocity to the lower left boundary:
 \begin{align}v = \begin{cases} 
      \frac{t}{t_0}v_0\quad t \leq t_0, \\
      v_0 \quad t > t_0, \\
   \end{cases}
\end{align}
with $v_0 = 16.5$ m/s and $t_0 = 1$ $\mu$s. The material properties are taken from \cite{borden2012phase}: $\rho = 8000$ kg/m$^3$, $E = 190$ GPa, $\nu = 0.3$, and $G_\text{C} = 2.213 \times 10^4$ J/m$^{2}$. We use $k = 1 \times 10^{-12}$ as the small constant used for preventing $\mathcal{W}^{e,+}$ from vanishing. We model the initial crack as an explicit discontinuity that resembles a sharp wedge. We use $\ell = 3.9\times 10^{-4}$ m (equivalently $\epsilon = 1.95\times10^{-4}$ m as in \cite{borden2012phase}) and $t = 0.04$ $\mu$s. We refine elements around where the crack is expected to propagate and ensure the element size within the refined region satisfies $\delta < \ell/2$. Figs. \ref{kalthoffdata}(b) and \ref{kalthoffdata}(c) show the temporal evolution of the total elastic energy and dissipated energy obtained from our simulator, respectively. Our results agree well with those extracted from \cite{borden2012phase} that considered multiple element sizes. Fig. \ref{kalthoffgraph} shows the temporal evolution of crack trajectory for four different time instants, all of which agree qualitatively with measurements from experiments \cite{kalthoff2000modes} and simulations \cite{borden2012phase}.

\section{Introduction to our crack tip tracking algorithm}
\begin{figure}[h]
\centering
\includegraphics[width=0.8\linewidth]{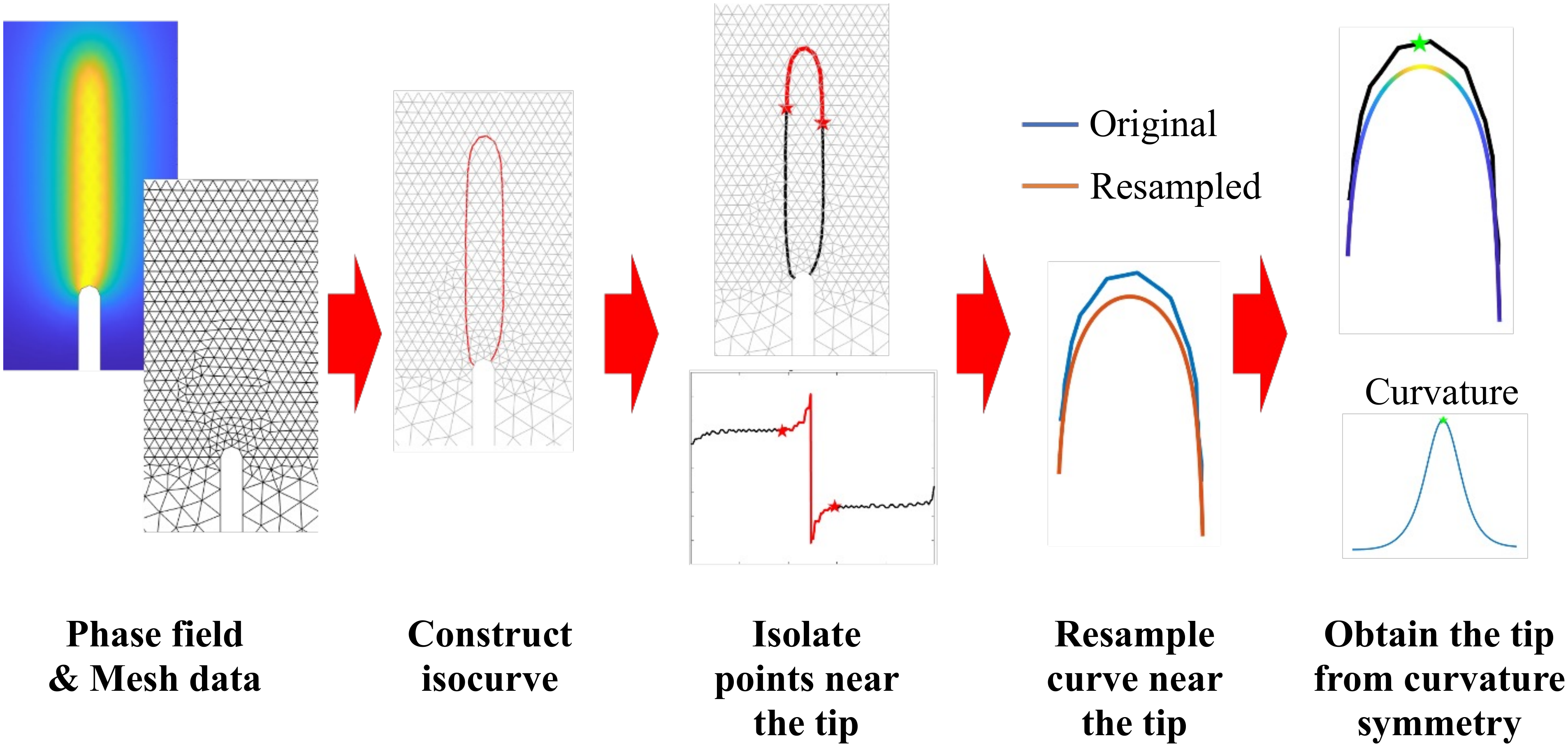}
\caption{A schematic demonstrating our crack tip tracking method.}
\label{cracktipidentify}
\end{figure}

We identify the boundary of a crack using a user-defined phase field value $\phi_\text{c}$. In this work, we pick $\phi_\text{c} = 0.85$. This algorithm finds what can be considered the tip of the boundary (i.e., the crack tip) in four steps (see Fig. \ref{cracktipidentify}): starting from the phase field and mesh data at a particular time step, it first reconstructs the iso-curve, isolates points near the tip, resamples these points, then computes the tip by looking for symmetries in the curvature. The algorithm is efficient and parallelizable, as multiple times teps can be analyzed at once.

The algorithm first constructs the iso-curve from the phase field and mesh data of a given time step. Here, the iso-curve is represented by a list of ordered pairs, $[(x_1,y_1), (x_2, y_2), ... , (x_n,y_n)  ]$, where each ordered pair represents a point in which the isocurve intersects with an element edge. The algorithm computes this list by looping through all elements and stopping when encountering an element (denoted as $\#n$) whose nodal phase field values $\phi_{n1}, \phi_{n2}, \phi_{n3}$ satisfy $\phi_{ni} < \phi_\text{c} < \phi_{nj}$ for at least one edge of that element. Then, starting from the edge $(n_i, n_j)$, it uses knowledge of mesh connectivity to look for the next edge where $\phi_{ni} < \phi_\text{c} < \phi_{nj}$. It stores the $(x, y)$ location of where $\phi = \phi_\text{c}$ on an edge $(n_i, n_j)$ as the next element in the list $[(x_1,y_1), (x_2, y_2), ... , (x_n,y_n)  ]$. The algorithm terminates when no more adjacent edges $\phi_{ni} < \phi_\text{c} < \phi_{nj}$ can be found. It then travels in reverse to sample the rest of the iso-curve.

From the list $[(x_1,y_1), (x_2, y_2), ... , (x_n,y_n)  ]$, the algorithm moves to isolate points on the iso-curve that are near the approximate location of the tip. This step is necessary because the end goal is to obtain the tip location using symmetries in the curvature of the iso-curve. However, in particularly straight cracks (which have a uniform curvature of zero along the sides), or cracks with even curvature throughout, this method can easily misidentify the tip. Thus, points of $\phi = \phi_\text{c}$ near the crack tip are first found and isolated. This is done by finding the closest vector between two points that are within some user-defined threshold of being anti-parallel. Specifically, let $\mathbf{x}_1 = (x_i, y_i)$, $\mathbf{x}_2 = (x_j, y_j)$ be two points on the iso-curve where $i < j $. We wish to find the combination of $(i, j)$ such that the quantity $j-i$ is as small as possible, and that $\frac{\mathbf{x}_1 \cdot \mathbf{x}_2}{|\mathbf{x}_1||\mathbf{x}_2|} \approx -1$. In practice, we simply stop the algorithm when $\frac{\mathbf{x}_1 \cdot \mathbf{x}_2}{|\mathbf{x}_1||\mathbf{x}_2|}$ is within some range centered at $-1$. Following this step, an approximate envelope consisting of the tip can be identified by a new list containing a reduced number of points $[(x_i,y_i), (x_{i+1}, y_{i+1}), ... , (x_{j-1},y_{j-1}), (x_j,y_j)  ]$. From this new list the points are resampled with a greater density using linear interpolation and a Gaussian smoothing process (to remove discontinuities in the curvature). Denoting the resampled curve to be $[(x_1^R, y_1^R) ,..., (x_{n_R}^R, y_{n_R}^R)]$, where $n_R$ is the total number of points created in the resampling process. Then, a simple curvature calculation is performed on this curve using numerical differentiation. More specifically, since the points of the iso-curve are in an ordered list, the curvature $\kappa$ at any point $i$ can be found by

\begin{align}
\kappa_i = \frac{\left[  (x_{i+1}-x_{i-1})^2+(y_{i+1}-y_{i-1})^2\right]^{3/2} }{\left|  (x_{i+1}-x_{i-1})(y_{i+2}-2y_i+y_{i-2}) - (x_{i+2}-2x_i+x_{i-2})(y_{i+1}-y_{i-1})    \right| },
\end{align}
which is essentially the numerical formulation of the curvature of a pair of parameterized functions in Cartesian coordinates:

\begin{align}
\kappa = \frac{\left[   (x')^2+(y')^2  \right]^{3/2}  }{\left|  x'y''-x''y' \right|}.
\end{align}
This calculation results in a list of curvatures $\left( \kappa_3, \kappa_4, ... , \kappa_{n_R-2}  \right)$, and the algorithm proceeds to look for the point $\kappa_i$ where the curvature plot is most symmetrical within some window $l$. This is done by summing the quantity $(\kappa_{i-k}-\kappa_{i+k})^2$ where $k = 1,2,...,l$. Essentially, we compute the difference between a point with an indicial distance $k$ on the left-hand side of $i$, and a point with an indicial distance $k$ on the right-hand side of $i$. We square this difference, and we sum this value over all possible values of $k$ from $1$ to $l$. This means computing:

\begin{align}
\text{Error associated with point}\,\,  i = \sum_{k = 1}^{l} (\kappa_{i-k} - \kappa_{i+k})^2.
\end{align}
The lower this error, the more symmetrical the curve is around the point $i$. Since we need to use a window of size $l$ to calculate this error, we simply do not consider points within an indicial distance of $l$ from the end of the resampled curve to avoid issues with this calculation. Then, the point with the lowest error is denoted as the crack tip for the considered time step, and the actual location of the tip is found by the identical index $i+2$ in the non-smoothed version of the resampled curve. We add this value of $2$ because the curvature calculation removed two entries from the indices.



\bibliographystyle{unsrt}
\bibliography{References}
\end{document}